\documentclass{ws-ijgmmp}

\usepackage{xcolor}
\usepackage{titlesec}
\newcommand{\tred}{\textcolor{black}}
\newcommand{\tteal}{\textcolor{black}}

\begin{document}

\markboth{Shaswata Chowdhury}
{Modified gravity in anisotropic stellar and substellar objects}

%
\catchline{}{}{}{}{}
%

\title{\uppercase{Anisotropies and modified gravity theories in stellar and substellar objects}
}

\author{SHASWATA CHOWDHURY}

\address{Department of Physics,
	Indian Institute of Technology Kanpur,\\
	Kanpur, 208016, India.\\
\email{shaswata@iitk.ac.in}}



\maketitle

\begin{history}
\received{(Day Month Year)}
\revised{(Day Month Year)}
\end{history}

\begin{abstract}
	In several classes of modified gravity theories, extra degrees of freedom are not completely screened in the interiors of stellar and substellar objects. In such theories, the hydrostatic equilibrium condition inside these objects is altered. Moreover, the interior structures of these objects might have a small pressure anisotropy induced by several physical phenomena, including rotation and magnetic fields. All these effects, both individually and collectively, induce changes in predicted stellar observables. Such changes have an impact on different phases of the stellar life cycle, starting from its birth to its death, covering almost all the branches of the Hertzsprung-Russell diagram. The aim of this work is to systematically review the current literature on the topic. We discuss the main results and constraints obtained on a class of modified gravity theories.
\end{abstract}


\section{Introduction}	

\subsection{Modified gravity theories}
The General Theory of Relativity (GR) \cite{Weinberg1972,Poisson2004}, formulated by Einstein more than a century ago, still stands as
the most successful theory of gravitational interactions. Its predictions have been validated by several precision tests, ranging from submillimeter \cite{SubmillimeterPRL,SubmillimeterPRD} to solar system scales (see \cite{GR_Experiment_Review,SubmillimeterReview} for reviews). It has been further established through the recent imaging of supermassive black holes \cite{BHimageM87,BHimageSagA}, as well as the detection of gravitational waves from compact object mergers \cite{GWreview}. However, despite its long standing success, it has been realized recently that GR possibly needs modifications, by incorporating extra degrees of freedom to the conventional Einstein-Hilbert action (for instance, via higher-order corrections in curvature scalars, and/or extra fields). The motivations behind developing such alternative theories of gravity are plenty \cite{ModGR_Hints}, the primary ones being the existence of mathematical singularities in GR \cite{SingularityGR} (which indicates a breakdown of the theory), the need for reconciling quantum mechanics with GR \cite{EffectiveGR}, and the issues relating to the observed cosmic acceleration and the cosmological constant \tred{$\Lambda$}. Furthermore, there have been observations on solar system scales \cite{ModGravSolarSystem} as well as galactic scales \cite{ModGravGalactic}, which hint at possible deviations from GR; however, such claims are not fully established at this point. In the literature there exists several classes of modified theories of gravity like $f(R)$ gravity \cite{fR_review}, Born-Infeld inspired gravity \cite{BI_review}, scalar-tensor theories (STTs) \cite{STT_review}, to name a few. These are called metric theories of gravity, where, like in GR, the metric is the only independent field with respect to which the corresponding action of the theory is varied, to obtain the gravitational field equations. However, complementing these metric theories of gravity there are metric-affine Palatini generalizations of these (e.g., Palatini $f(R)$ \cite{PalatinfR_reviewOlmo}, Eddington-inspired Born-Infeld (EiBI) gravity \cite{EiBI}, Palatini STT \cite{PalatiniSTT}), where both the metric and the connection are regarded as independent fields, with the matter field coupling only to the metric (and perhaps to the metric compatible Levi-Civita connection) \cite{PalatinfR_reviewOlmo}. Connections are geometrical objects on smooth manifolds, which help in defining differentiation of sections of vector bundles \cite{JohnMLee}, i.e., they define how the bases of vector spaces at {\it infinitesimally close} neighbouring points are related (or {\it connected}) \cite{Blau}. Purely metric-affine theories on the other hand allows for the coupling of matter to any general connection besides the metric itself \cite{PureMetricAffine}. \tteal{Different classes of black holes and their properties have been studied in different modified gravity theories like $f(R)$ gravity \cite{NashedBHfR,NashedRotBHfR,NashedChargedBHfR1,NashedChargedBHfR2}, mimetic gravity \cite{NashedChargedRotBHmimetic}, and teleparallel gravity \cite{NashedEnergyMomTele}. Different kinds of space-time solutions, like Reissner-Nordstr\"{o}m and Kerr-Newman, have also been investigated in different modified gravity theories \cite{NashedRNTetrad,NashedRNTele,NashedKNTele}. Effects of modified gravity on the cosmic inflationary model can be found in \cite{NashedInflationfT}.}

\tred{All the viable modified gravity theories in the literature should in principle conform with the observations on large scales in the late-time universe -- the accelerated cosmic expansion  and galactic rotation curves. In GR, the cosmic expansion is attributed to the dark energy associated with the cosmological constant $\Lambda$, and the galactic rotation curve is attributed to dark matter. Both these elusive entities -- dark energy and dark matter, suffer from severe criticism and challenges from both observational and theoretical perspectives. Therefore, it is but natural to consider modified gravity theories, which can explain the late-time universe without any 'dark' quantities. For example, the STTs and $f(R)$ gravity theories admit self-accelerating solution\footnote{\tred{By self-acceleration, we mean acceleration of the space-time due to modified gravity and not due to the energy associated with cosmological constant $\Lambda$.}} (see \cite{CrisostomiKoyamaSelfAcc,CapozzielloSelfAccfR}) and are alternatives to the dark matter model as well (see \cite{MotaDM2,BekensteinMilgromDM1,CapozzielloDM}).}
Recently however, the Normalized Additional Velocity (NAV) approach to study galaxy rotation curves indicates that the Palatini $f(R)$ and EiBI might not be suitable candidates to replace dark matter in galaxies \tred{\cite{WojnarNAV,WojnarNAV2}}. \tred{Other interesting phenomena in the late-time universe, like large-scale structure formation \cite{MotaLargeStructure} and halo abundances, have been extensively studied in the modified gravity theories through N-body simulations and compared with those derived from $\Lambda {\rm CDM}$ model\footnote{\tred{The $\Lambda {\rm CDM}$ model corresponds to one of the most successful cosmological models in GR with dark energy ($\Lambda$) and cold dark matter (CDM).}} in GR. The authors found some signatures of modified gravity theories in these studies. However, it was later observed that the predictions from modified gravity theories conform with those from the $\Lambda{\rm CDM}$ model when massive neutrinos are taken into account \cite{MotaNeutrino}. So, the natural question to ask at this stage is, then, what is the smoking gun test for modified gravity theories? The velocity profile of galaxies within a cluster \cite{MotaVelProfileSmokingGun}, derived from modified gravity theories, shows deviations when compared to those derived from the GR models. This deviation is, in particular, attributed to the screening mechanism, which is an integral part of modified gravity theories. Therefore, it serves as a smoking gun test for modified gravity theories. We will discuss more about the screening mechanism shortly, but the details about the late-time universe in modified gravity theories are beyond the scope of this review.}

STTs, a class of which arises from incorporating scalar fields in the Einstein–Hilbert action, are one of the most popular and successful {\it avatars} of modified gravity theories. The initial work on STTs in the context of modified gravity was carried out in \cite{Sotiriou(2006)}. Over the past decade, substantial research has been conducted on STTs, their implications in cosmology \cite{ModGravDarkEnergy1,ModGravDarkEnergy2,ModGravDarkEnergy3} and various constraints on such theories \cite{TestsOfSTT} have been obtained. The most general theories of a scalar field coupled to gravity, which are ghost-free, constitute the Horndeski theories \cite{Horndeski}. By ghost-free theories we mean those that are free from Ostrogradsky instabilities \cite{Ostrogradsky1,Ostrogradsky2}, which result from non-degenerate Lagrangians containing second and higher order derivatives in the scalar field. The Horndeski theories are further generalised to the beyond Horndeski class of degenerate higher order scalar-tensor (DHOST) theories \cite{DHOST1,DHOST_review2}. All the STTs discussed in the literature are usually formulated in the Jordan frame \cite{ModGravDarkEnergy1}, where "frame" refers to a set of physical variables \cite{Sotiriou(2006)}. In the Jordan frame, the matter action includes only minimal coupling of the matter fields with the metric. Therefore, the energy-momentum tensor is covariantly conserved and test particles travel along geodesics of the metric, thereby satisfying Einstein's Equivalence Principle (EEP). The Einstein frame, on the other hand, is related to the Jordan frame through a conformal transformation \tred{(see \cite{EquivJordanEinstein1})}, such that the matter action includes non-minimal couplings of the matter fields with the metric and possibly scalar fields as well, which violates the EEP. Hence, Jordan frame is usually deemed to be the physical one. Although one can perform the mathematical calculations in the Einstein frame, one needs to return to the Jordan frame. \tred{For debates related to which frame is more suitable for the description of gravity, see a discussion in \cite{EquivJordanEinstein2}.}

Any scalar-tensor theory, which is an alternative to the dark energy model, should give rise to a modification in the gravitational interactions at cosmological scales. However, such modifications mediated by the scalar field must be screened at small scales where GR has been validated to high accuracy. The Vainshtein mechanism \cite{Vainshtein} is one of the most efficient screening mechanisms, in which GR is recovered at small scales, via non-linear effects (see, e.g. \cite{VainshteinReview1,VainshteinReview2} for reviews). However, it has been shown by \cite{PartialBreakingVainshtein} that in DHOST theories, the Vainshtein mechanism breaks inside matter densities, i.e., the effect of such modified gravity theories is not fully screened inside stellar and substellar objects, as well as astronomical objects like galaxies, and galaxy clusters \cite{DHOSTconstraintGalaxy}. An interesting consequence of this is that in the low energy (Newtonian) limit, the pressure balance equation inside stellar and substellar objects is modified. The pressure balance equation is an essential component in deriving the analytical formulas for stellar and substellar observables. Hence, one can constrain this class of modified gravity theories from observational data. Several studies have been reported in this domain in recent years, starting from the pioneering research conducted by \cite{KoyamaSakstein} (see, e.g., \cite{SaitoJCAP,SaksteinPRL,SaksteinPRD,JainWD,BabichevKoyama,SaksteinStrongFieldTest,SaltasWD,Chowdhury1,SaltasHelios,Pritam1,SaksteinAstroTests}). To obtain the recent information on the implications of modification of gravity in stellar objects, see \cite{OlmoDiegoWojnarReview,WojnarReviewPalatini}. 

In fact, after the nearly simultaneous detection of gravitational waves from a pair of inspiralling neutron stars GW170817 \cite{GW170817Paper} and its electromagnetic counterpart GRB 170817A \cite{GRB170817Paper}, several STTs have been tightly constrained \cite{STTafterGW1,STTafterGW2,STTafterGW3,STTafterGW4,STTafterGW5,STTafterGW6,STTafterGW7,STTafterGW8,STTafterGW9}. Nevertheless, there have been efforts invested by the gravity community to carefully analyze the validity of such constraints imposed on several modified gravity models \cite{ValidityOfGWonSTT}. There have also been attempts to revive some of the heavily constrained theories as well; for example, \cite{SaveHorndeskiTeleparallel} tries to revive the class of Horndeski theories using teleparallel gravity \cite{TeleparallelGR}.

Even if general relativity turns out to be the true theory of gravity in the end, it is worthwhile to study modified gravity theories because of two primary reasons. Firstly, to test gravity, it is necessary to understand the predictions of theories other than general relativity. Secondly, consistent alterations of gravity help us gain a deeper insight into general relativity and gravity as a whole. Indeed, experiments need to be used as a guide to confine the range of possibilities while constructing an alternative theory of gravity. Among the numerous scenarios suggested in the literature to test gravitational interactions beyond GR, the stellar structure models represent an effective probe for detecting eventual deviations from GR predictions.

\subsection{Stellar and substellar physics}
\label{SecStellarPhysics}

Stellar objects, commonly called stars, are self-gravitating astrophysical objects, where different types of nuclear reactions take place at different points of time. The initial nuclear reaction in a star converts the lightest element hydrogen into helium. Such a phase is called the main sequence phase, where the central hydrogen is converted into helium through either proton-proton (pp) chain reactions or the Carbon-Nitrogen-Oxygen (CNO) cycle. Depending upon the mass and age of the star, subsequent nuclear reactions follow after the main sequence phase, synthesizing heavier elements. The different classes of stars like the main sequence stars (MSS)s, sub-giants, red-giants, white dwarf (WD) stars, and many more, represent the different phases of the stellar life-cycle. Substellar objects on the other hand, are the astrophysical objects which never attain the conditions favourable for the initial nuclear reaction within. Such objects have masses less than a particular minimum value called the minimum mass of hydrogen burning $M_{{\rm mmhb}}$ $\sim 0.08M_{\odot}$. This is also sometimes referred to as the minimum main sequence mass. Brown dwarfs (BD)s are the examples of such substellar objects. Stars in the mass range $\sim 0.08 M_{\odot} - 0.3 M_{\odot}$ are called very low mass (VLM) stars. One collectively calls the VLM stars and brown dwarfs as VLM objects. Stars in the mass range $\sim 0.3 M_{\odot} - 1.4 M_{\odot}$ are called low mass stars, while those within the range $\sim 1.4 M_{\odot} - 8 M_{\odot}$ are called intermediate mass stars. Stars having mass higher than $\sim 8 M_{\odot}$ are called high mass stars. The Hertzsprung-Russell (HR) diagram catalogues the luminosity and effective temperature of different classes of stars, from observations. It reveals the different phases of stellar evolution. 
\subsubsection{Minimum mass of hydrogen burning}
The self-gravity of a newly formed VLM object causes it to contract in size throughout the early phases of its existence. In the process, the decrease in its gravitational potential energy is manifested as the energy radiated from its surface, which is referred to as the surface luminosity, or simply luminosity. The contraction leads to an initial increase in its central density and temperature, while the surface luminosity keeps decreasing. At some stage, if the attained thermal energy of the nucleons is sufficient enough for them to quantum-tunnel through the coulomb potential energy barrier of a hydrogen nucleus, then the thermonuclear reaction kick-starts. However, in VLM objects, the thermal energy is not high enough to trigger a full pp chain reaction; rather a truncated pp chain reaction takes place \cite{BurrowsLiebert}. The energy generated from this thermonuclear reaction is referred to as the hydrogen-burning luminosity. With further contraction of the object, the hydrogen burning luminosity increases. At some stage, if the amount of energy produced from thermonuclear reaction is balanced by the energy escaping the surface, the stellar object is said to attain stable/sustained hydrogen burning. The object does not undergo further contraction, and it is said to become a MSS. However, if the object's initial mass, after formation, is smaller than $M_{{\rm mmhb}}$, then significant degeneracy develops before the onset of stable hydrogen burning. The temperature starts falling with further contraction. This is due to the fact that in order to accommodate a large number of degenerate electrons in a small volume, a part of the thermal energy is used up. Such a situation does not allow the stellar object to attain stable hydrogen-burning condition. 
Therefore, all stellar objects above $M_{{\rm mmhb}}$ reach the main sequence after its formation, while those below it can never reach the main sequence and are thus deemed to be failed stars; such substellar objects are called BDs. BDs were theoretically first predicted by \cite{ShivSKumar,HayashiNakano} and the analytic models for such VLM objects were provided by \cite{BurrowsLiebert,DAntona1,Burrows1989}. BDs were first observed by \cite{Rebolo1995,Nakajima1995}. After their discovery, all further developments on BDs have been well documented in \cite{ChabrierBaraffe2000, Basri2000, Burrows2001, Rebolo2000,Burrows1997,ChabrierBaraffe1997,DAntona1997}. More recent reviews can be found in \cite{Joergens2014, Allard2012, Chabrier2014, MarleyRobinson}.

\subsubsection{Evolution of low and intermediate mass stars away from ZAMS}
\begin{figure}[h]
	\centering
	\begin{minipage}[t]{0.40\linewidth}
		\begin{center}
			\centerline{\includegraphics[scale=0.22]{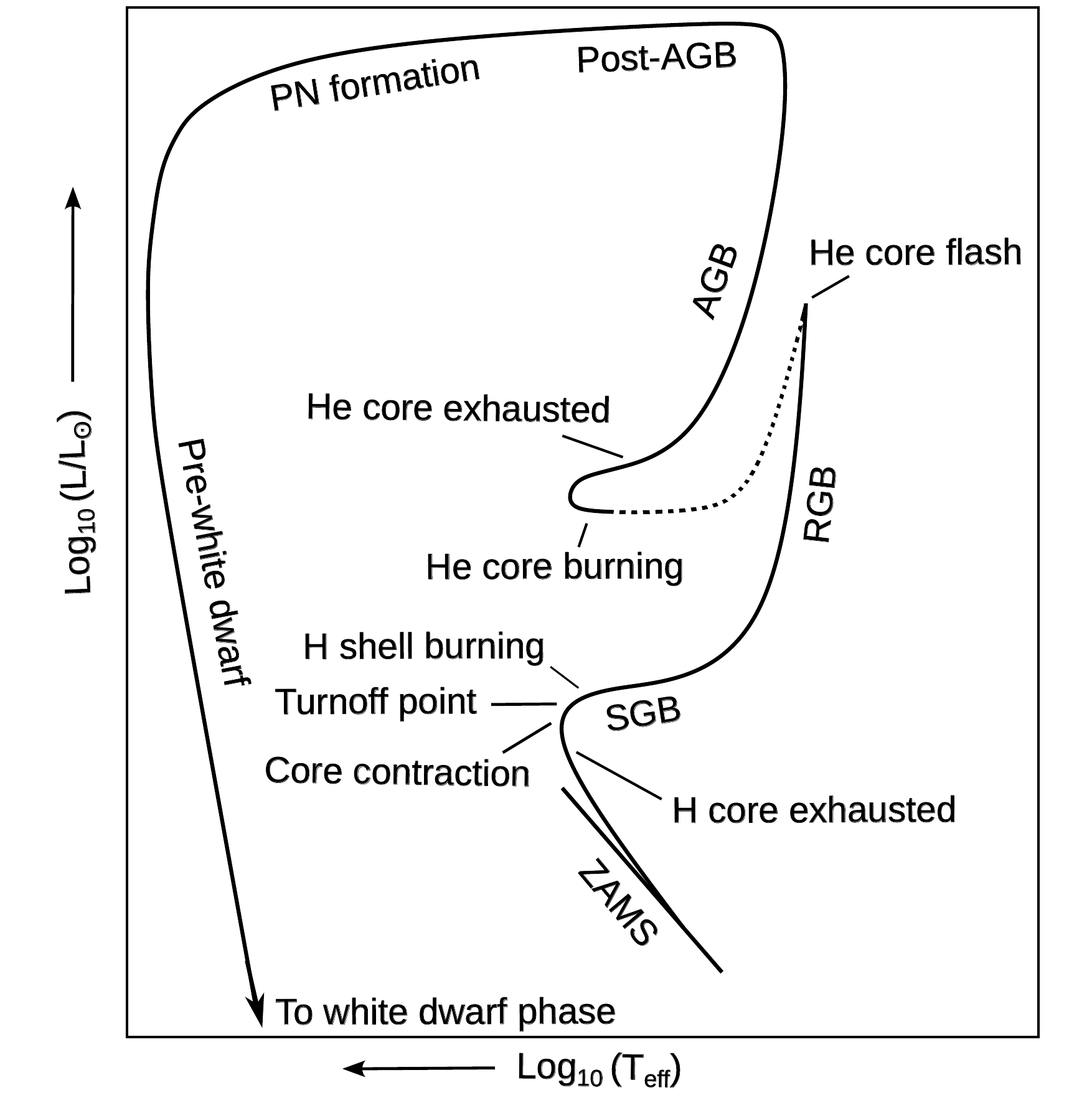}}
			\caption{A schematic diagram of the evolution of a low-mass star from the ZAMS up to the white dwarf phase. $L$ corresponds to luminosity and $T_{\rm eff}$ is effective temperature. This figure is inspired by \protect\cite{CarrollOstlie}.}
			\label{FigHR_low}
		\end{center}
	\end{minipage}%
	\hspace{0.9cm}
	\begin{minipage}[t]{0.40\linewidth}
		\begin{center}
			\centerline{\includegraphics[scale=0.22]{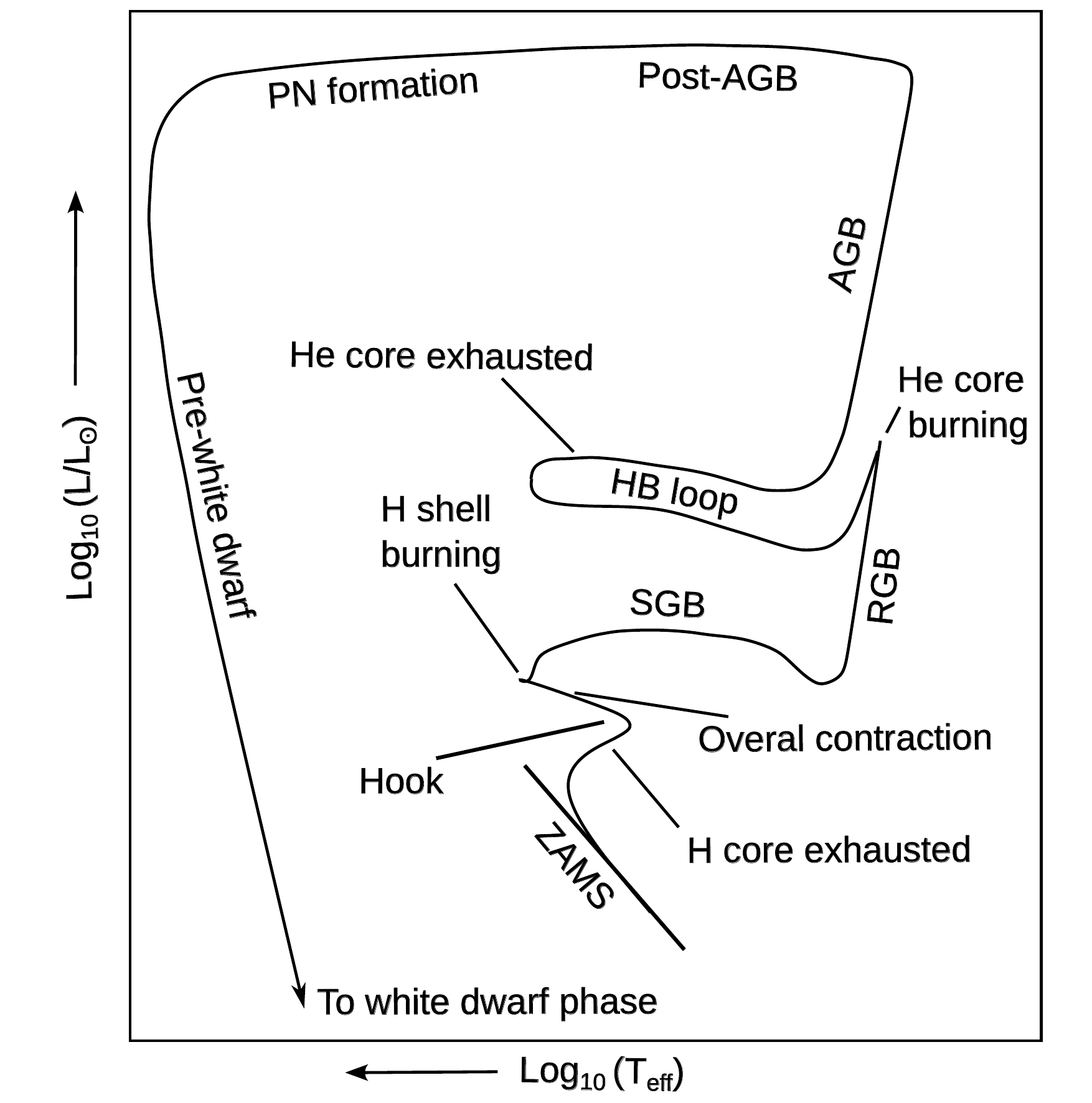}}
			\caption{A schematic diagram of the evolution of an intermediate-mass star from the ZAMS up to the white dwarf phase. $L$ corresponds to luminosity and $T_{\rm eff}$ is effective temperature. This figure is inspired by \protect\cite{CarrollOstlie}.}
			\label{FigHR_intermediate}
		\end{center}
	\end{minipage}
\end{figure}

The stage when a star just attains the main-sequence phase is called the zero-age-main-sequence (ZAMS). Depending upon the initial mass of a star, its evolution away from the ZAMS goes through a sequence of different distinctive phases. See for example Fig.\ref{FigHR_low} and Fig.\ref{FigHR_intermediate}, \cite{CarrollOstlie}. The various phases of evolution labeled in the figures are ZAMS, sub-giant branch (SGB), red giant branch (RGB), asymptotic giant branch (AGB), post-asymptotic giant branch (Post-AGB), planetary nebula formation (PN formation), pre-white dwarf phase leading to white dwarf phase, and the horizontal branch (HB) loop.
We see that the evolution of low mass and intermediate mass stars share a lot of common features, apart from the hook formation post hydrogen core exhaustion, and the HB loop post helium core-burning in the intermediate mass stars. The reader is referred to the excellent textbook on stellar physics \cite{CarrollOstlie} for a more detailed analysis; here, we only give a brief overview of the stellar evolution of low and intermediate-mass stars, closely following \cite{CarrollOstlie}.

For low mass MSSs, hydrogen gets converted to helium through the pp chain reaction, where the nuclear reaction rate goes as $\rho X^2T_6^4$, with $\rho$ being the density, $X$ being the hydrogen mass fraction and $T_n$ in general being the temperature $T$ in units of $10^n$K. According to the ideal gas law, with an increase in the mean molecular weight of the central region due to helium production, the central temperature and density should also increase in order to counterbalance the gravitational pressure due to overlying layers of the star. The core, therefore, needs to compress and, in the process, releases gravitational potential energy. According to the virial theorem, half of this energy is used up in increasing the kinetic (thermal) energy of the gas and thus the temperature, while the other half is radiated away. The increase in density and temperature results in the increase in energy production -- and thus the overall stellar luminosity and effective temperature. The core contraction is accompanied by an expansion of the envelope, which leads to an overall increase in the stellar radius. This is called the ``mirror principle,'' which says that core contraction is accompanied by envelope expansion and vice versa. Although this is an interesting rule of thumb which is widely used in explaining a lot of stellar physics starting from the ZAMS to the supernovae, it is hard to justify from a single theory as of now. 

As the star continues to evolve along the main sequence, eventually its central hydrogen gets depleted, leaving behind an isothermal helium core \cite{Gamow}. The main-sequence phase thus comes to a halt. Hydrogen burning in the core during the main-sequence, however, raises the gas temperature enough to keep hydrogen burning in a thick shell surrounding the core. This phase is referred to as the hydrogen shell-burning phase, post main-sequence. Since the hydrogen shell-burning occurs at a much higher temperature compared to that during core hydrogen burning, the energy generation is even more during this shell-burning phase. Although a portion of this energy is being used up in slow expansion of the stellar envelope -- thus increasing the stellar radius, a significant amount of it reaches the stellar surface. Therefore, the stellar luminosity increases, while the effective temperature starts decreasing slightly due to the increase in stellar radius. This is referred to as the turnoff point in the HR diagram. The stellar models for stars in their turnoff points were first provided by \cite{HS}.

The hydrogen-burning shell keeps adding mass to the helium core, post main-sequence. For intermediate mass stars, this phase continues until the point when the core mass fraction attains a certain maximum and can no longer support the pressure of the overlying envelope. This maximum core mass fraction is called the Sch\"{o}nberg-Chandrasekhar (SC) limit \cite{HenChan,SC}, and it is a function of the ratio $\alpha$ of the mean molecular weight of the core $\mu_c$ to that of envelope $\mu_e$ (see \cite{CoxGiuli,Kippenhahn}). After the SC limit is reached, the hydrogen-depleted core contracts rapidly on a Kelvin-Helmholtz timescale, and thereby releasing gravitational potential energy. This expands the stellar envelope, resulting in a decrease in the effective temperature. This phase of very rapid redward evolution, known as the subgiant branch (SGB), on the HR diagram results in the Hertzsprung gap. However for low mass stars, the core develops degeneracy pressure which can counterbalance the pressure of the overlying envelope, even for core mass fractions exceeding the SC limit. On the other hand, for higher mass stars, after the central hydrogen-burning ceases in the core, the shell burning does not start immediately. Instead the entire star contracts on a Kelvin-Helmholtz timescale. Such a contraction releases gravitational potential energy, which causes the luminosity to increase slightly and since the overall radius of the star decreases, the effective temperature increases. This is the ``hook formation'' in the evolution of higher mass stars in the HR diagram.

As the stellar envelope expands and the effective temperature decreases, the more stable $H^-$ ions start dominating the photospheric region. These ions contribute to the higher absorption of outbound photons emanating from the interior, thus resulting in increased photospheric opacity. As radiation becomes less effective in transporting energy in the high-opacity photospheric region, convection sets in near the stellar surface. Since convection is a highly efficient mechanism of energy transport, the luminosity rises rapidly, leading to the nearly vertical ascent along the RGB of the HR diagram.

By the time when an intermediate mass star reaches the tip of the RGB, the central temperature and density attains sufficiently high values to trigger helium burning reaction in the stellar core, besides the shell hydrogen-burning, which is still the dominant source of stellar luminosity. This helium burning converts helium into heavier elements like carbon and oxygen. It takes place through the triple alpha process, with an energy generation rate that goes as $\rho^2 Y^3 T_8^{41}$, with $Y$ being the helium mass fraction. The strongly temperature dependent helium burning reaction generates core pressure that exceeds the gravitational pressure of the overlying envelope. This leads the core to expand and thus push the hydrogen-burning shell outward, resulting in a decrease of the shell temperature. Therefore the rate of energy generation from the hydrogen shell-burning decreases, causing the overall stellar luminosity to decrease abruptly. According to the mirror principle, the envelope contracts, leading to the decrease in the stellar radius and an increase in effective temperature. However for low mass stars, an interesting thing happens at the tip of its RGB. With the contraction of the helium core along the RGB, degeneracy develops within. Such a degeneracy pressure is the dominant source of outward pressure for the core, in the absence of any thermonuclear reactions inside the core. With further contraction of the core, when the central temperature and density becomes high enough, helium burning reaction starts. At this point the degeneracy pressure still dominates the thermal pressure due to nuclear reaction. The highly temperature dependent helium burning reaction increases the central temperature which in turn leads to higher energy generation. However, the degeneracy pressure being temperature insensitive, remains the same. With this positive feedback mechanism, the temperature and thus the thermal pressure keeps increasing enormously, until the point when the thermal pressure dominates over the degeneracy pressure. At this point the degeneracy of the core is lifted and the tremendous amount of thermal energy is expelled out in a few seconds. This appears as a flash and is called the helium core flash. The luminosity of such an event reaches $\sim 10^{11}L_{\odot}$, which is comparable to that of an entire galaxy.

Following the tip of the RGB, both the low and intermediate mass stars undergo core expansion accompanied by contraction of the envelope. As mentioned above, this initially leads to an abrupt descent in the corresponding evolutionary track. However, since the expanding core and the contracting envelope compresses the hydrogen-burning shell, the temperature and density of the shell increases in the process. Therefore a time comes, when the shell temperature and density are high enough to result in a slow increase in the stellar luminosity and effective temperature. This horizontal evolution marks the blueward portion of the HB loop. Similar to the main-sequence phase, here also the central helium gets depleted after some time, which results in the contraction of the carbon-oxygen (CO) core and rapid redward evolution along the HB loop, analogous to the SGB. 

With the contraction of the CO core after depletion of the central helium, the central temperature and density rises back to the point where now helium starts burning in a shell surrounding the CO core. The helium shell-burning takes place at a much higher temperature compared to that during core helium burning. Therefore the energy liberated and thus the luminosity is much higher during this helium shell-burning phase. The effective temperature however, decreases owing to the expansion of the envelope which accompanies the core contraction by the mirror principle. This marks the nearly vertical AGB phase which is analogous to the shell hydrogen-burning RGB. The helium shell-burning keeps adding mass to its CO core. During the AGB phase, the star loses mass from its extended envelope at a rapid rate, through stellar winds \cite{HabingOlofAGB,RamstedtEtalAGB}.

The remainder of the stellar envelope is ejected during the subsequent last phase of mass loss, exposing the cinders created by the extensive nuclear reactions inside the star. The ejected stellar envelope forms the planetary nebula, while the exposed hot central CO core, surrounded by thin layer of remaining hydrogen and helium, eventually cools to become a white dwarf star. The maximum mass of WD being theoretically predicted as $1.4M_{\odot}$ by S. Chandrasekhar, is known as the Chandrasekhar limit.

\subsection{Anisotropies}
So far, all of the stellar structure and evolution that we introduced in the previous section Sec.\ref{SecStellarPhysics} were for isotropic situations where the stellar object possesses spherical symmetry. Although such ideal isotropic models can explain the essential physics to a large extent, it is well known that stellar objects need not be isotropic in reality. In such an anisotropic situation, the radial and tangential stresses are not equal. Sometimes it is also referred to as stellar pressure anisotropy. Over the past few decades, anisotropic stars have been the subject of intense research \cite{Herrera1,Herrera2,Herrera3}. There are a number of possible reasons for anisotropy in stellar structures within a Newtonian framework. These include stellar rotation and the presence of magnetic fields within the stellar interior \cite{Spruit}, both of which have the potential to oblate (or prolate) the stellar structure. Although observations suggest that the magnitudes of such distortions may be low, studying these anisotropies is essential for a full understanding of stellar dynamics.

\subsubsection{Rotation}
Stellar and substellar objects might have slow as well as rapid rotations; rotating polytropes have been studied since the works of \cite{ChandraRot1933} (see \cite{Monaghan,Kovetz,Cook,Kong,Chowdhury3,Ballot,Yoshida,Reina}). Stellar rotation periods of $\sim 17$ minutes \cite{Route} to $\sim 3$ days \cite{Joergens2003} have been observed for brown dwarfs. Rotation periods of white dwarfs range from $\sim 25$ seconds \cite{WhiteDwarfRotObs1} to $\sim 2$ days \cite{WhiteDwarfRotObs2}. Rotation periods of main-sequence stars range from $1$ to $30$ days in F to M spectral types \cite{MSRotObs}.
Giant planets have been reported to have rotation periods of $\sim 15$
hours \cite{GiantPlanetRotObs}. Therefore stellar rotation is an important issue that might give rise to new theoretical possibilities and can explain exotic events in astronomy. For example the consideration of differential rotation in white dwarfs \cite{Ghoshwheeler} is crucial for the theoretical understanding of the observed super-Chandrasekhar progenitor white dwarfs, in the mass range of $2.1-2.8 M_{\odot}$ \cite{Scalzo,Hachisu}. The consideration of rapid rotation in VLM objects \cite{Chowdhury3}
predicts overmassive BDs, having masses higher than $M_{{\rm mmhb}}$. Studying rapid rotations need fully numerical treatment \cite{Chowdhury3}, while slow rotations can be treated analytically \cite{ChandraRot1933, Chowdhury5}.  
\subsubsection{Magnetic fields}
Recently, in \cite{Berdyugina2017}, it was discovered that low-density brown dwarfs exhibit magnetic activity (magnetic fields $\sim10^3-10^4$ G at the surface). The surface magnetic fields of post main sequence stars are also $\sim 10^4$ G (see, e.g., \cite{QuentinTout2018}). The magnetic fields of white dwarfs can be maximally $\sim 10^9$ G, but those of magnetic neutron stars can reach up to $10^{15}$ G. The existence of magnetic fields induce pressure anisotropy because rotational symmetry is broken \cite{Ferrer2010}. Both \cite{Ferrer2010} and a trio of studies by \cite{CanutoChiuA,CanutoChiuB,CanutoChiuC} provide analytical determinations of this anisotropy in the presence of strong magnetic fields. However, if the magnetic fields are small then the induced stellar pressure anisotropy is small enough to approximate spherical symmetry. In such a situation, the stellar pressure anisotropy is modeled as in \cite{Chowdhury1,HeintzmannHillebrandt}. Similar to stellar rotation, magnetic fields incorporated in theoretical models gives rise to new predictions. For example, strongly magnetized white dwarfs with central magnetic fields $\sim 10^{16}$G, were proposed by \cite{BaniDasPRL} to be possible candidates for the super-Chandrasekhar white dwarf progenitors. In this work, the authors considered the modified equation of state for a highly magnetized Fermi gas. Regarding the stability of such models under the consideration of Lorentz forces, follow up articles \cite{NityanandaKonar,BaniDasPRD} can be looked at. However, it was pointed out that the higher magnetic moment terms which appear in the pressure balance equation due to the Lorentz force might be approximated by an equivalent stellar pressure anisotropy term \cite{Chowdhury1}.

\subsection{Aim of this work}

\tred{Most of the earlier works on stellar structure in modified gravity theories were confined to compact objects like neutron stars, which are deemed to be the primary candidates for testing gravity in its strong field regime, owing to the availability of large amounts of observational data. However, the less compact stellar and substellar objects remained a poorly studied branch in modified gravity theories until the past few years, when the Newtonian limit of these modified theories has been significantly explored. In such studies, the anisotropic situation remains an interesting and more realistic aspect of stellar physics owing to different physical phenomena like rotation and magnetic fields.}

This work, \tred{therefore,} aims to review anisotropic stellar and substellar objects within the framework of a certain class of DHOST theories beyond Horndeski \tred{in the Newtonian limit}. Here we will discuss the predictions of such a framework on the stellar and substellar observables, as well as how one obtains bounds on the parameters characterizing modified gravity theories and anisotropies. We will elaborate on the status of this particular field of research after GW170817 and the possible future directions in this line of work.

Throughout the text, the terms {\it alternative} and {\it modified} in the context of theories of gravity will be used synonymously. Similarly, throughout the text, both {\it hydrostatic equilibrium condition} and {\it pressure balance equation} will have the same meaning. To avoid confusion, we shall refer to the isotropic Newtonian limit of GR as the "standard" case for the rest of this study. Moreover, we will adhere to the metric signature $(-,+,+,+)$ throughout the text.

The content of this paper is organized as follows. In Sec.\ref{SecDHOST}, the interesting features of DHOST theories inside astrophysical objects are reviewed starting from the galileon theory and the current status of such theories after GW170817 is elaborated. In Sec.\ref{SecModAnisotropy}, we review the results and constraints obtained by modeling stellar and substellar objects with local pressure anisotropies, within the framework of DHOST theories, retaining the approximation of spherical symmetry; the VLM objects, low and intermediate mass stars at different points of their evolution are considered. In Sec.\ref{SecRotation}, we describe interesting features in VLM objects when the approximation of spherical symmetry is relaxed by incorporating rapid rotation, in the Newtonian framework, and then provide an analytical formalism to include slow rotation in polytropic stars within the framework of any modified gravity theory in general. \tred{In Sec.\ref{SecStrongField}, we discuss the effects on stellar observables when strong field corrections are incorporated into the modified gravity theories. We also compare the constraints on modified gravity theories obtained from the neutron stars with those obtained from the less compact stellar and substellar objects, which constitute this review article.} We end the paper with discussion and future perspectives.

\section{DHOST theories inside stellar and substellar matter}
\label{SecDHOST}
In this section we will briefly review the DHOST theories from phenomenological perspective, along with some necessary mathematical details. We will highlight the essential features of the Vainshtein mechanism in STTs and its partial breaking inside astrophysical objects, within the framework of DHOST theories. Our notations closely follow the excellent review \cite{HorndeskiBeyondReview}.

\subsection{DHOST theories}

\subsubsection{Galileon theory}
In a fixed Minkowski background, a scalar-field theory whose equations of motion remain invariant under the transformation $\phi \rightarrow \phi + b_{\mu}x^{\mu} + c$, is called a Galileon theory. The associated scalar field, called the Galileon, is said to possess Galilean shift symmetry. The name is motivated by the analogous Galilean transformation in classical mechanics. In 4D, the most general ghost-free Galileon theory possessing appropriate non-linearities, is represented by the following Lagrangian containing only five terms \cite{GhostfreeGalileon}
\begin{equation}
\mathcal{L}_G = c_1\mathcal{L}_1 + c_2\mathcal{L}_2 + c_3\mathcal{L}_3 + c_4\mathcal{L}_4 + c_5\mathcal{L}_5~, 
\label{GalLag}
\end{equation}
where
\begin{align}
&\mathcal{L}_1 = \phi~,~~ \mathcal{L}_2 = X~,~~\mathcal{L}_3 = -X\Box\phi~,~~\mathcal{L}_4 = X\big[(\Box\phi)^2 - \partial_{\mu}\partial_{\nu}\phi\partial^{\mu}\partial^{\nu}\phi\big]~, \nonumber\\
&\mathcal{L}_5 = -\frac{1}{3}X\big[(\Box\phi)^3 - 3\Box\phi\partial_{\mu}\partial_{\nu}\phi\partial^{\mu}\partial^{\nu}\phi + 2\partial_{\mu}\partial_{\nu}\phi\partial^{\nu}\partial^{\lambda}\phi\partial_{\lambda}\partial^{\mu}\phi \big]~,
\end{align}
with $X=-\eta^{\mu\nu}\partial_{\mu}\phi\partial_{\nu}\phi/2$ and $c_1,....,c_5$ are constants. The requirement of {\it non-linearities} is related to Vainshtein screening which will be elaborated in Sec.\ref{SecVainshtein}.

\subsubsection{Covariant Galileon theory}
\label{SecCovGal}
One can now incorporate gravity and consider a covariant version of Eq.\eqref{GalLag}, using the principle of minimal coupling. However, such a naive covariant version is found to introduce higher derivatives in the field equations, thus leading to ghost-instabilities. Nevertheless, by adding suitable curvature-dependent terms, by hand in the Lagrangian, such higher derivative terms can be canceled out. Therefore the ghost-free covariant Galileon theory is given by the following Lagrangian \cite{CovariantGalileon}
\begin{equation}
\mathcal{L}_{CG} = c_1\tilde{\mathcal{L}}_1 + c_2\tilde{\mathcal{L}}_2 + c_3\tilde{\mathcal{L}}_3 + c_4\tilde{\mathcal{L}}_4 + c_5\tilde{\mathcal{L}}_5~, 
\label{CovGalLag}
\end{equation}
where
\begin{align}
&\tilde{\mathcal{L}}_1 = \phi~,~~\tilde{\mathcal{L}}_2 = X~,~~\tilde{\mathcal{L}}_3 = -X\Box\phi~,~~\tilde{\mathcal{L}}_4 = \frac{1}{2}X^2R + X\big[(\Box\phi)^2 - \nabla_{\mu}\nabla_{\nu}\phi\nabla^{\mu}\nabla^{\nu}\phi\big]~, \nonumber \\
&\tilde{\mathcal{L}}_5 = X^2G^{\mu\nu}\phi_{\mu\nu}-\frac{1}{3}X\big[(\Box\phi)^3 - 3\Box\phi\nabla_{\mu}\nabla_{\nu}\phi\nabla^{\mu}\nabla^{\nu}\phi + 2\nabla_{\mu}\nabla_{\nu}\phi\nabla^{\nu}\nabla^{\lambda}\phi\nabla_{\lambda}\nabla^{\mu}\phi \big]~,
\end{align}
where $R$ is the Ricci scalar, $G_{\mu\nu}$ is the Einstein tensor and now $X=-g^{\mu\nu}\nabla_{\mu}\phi\nabla_{\nu}\phi/2$. Here the first terms in $\tilde{\mathcal{L}}_4$ and $\tilde{\mathcal{L}}_5$ are introduced as ``counter terms" to cancel out higher derivatives from the field equations. The above Lagrangian thus yields second order field equations in both the scalar field as well as the metric. However the Galilean shift symmetry is broken in such a covariant Galileon theory \cite{HorndeskiBeyondReview}. From this point onwards, we will abide by the following compact notations: $\nabla_{\mu}\phi = \phi_{\mu}$, $\nabla_{\mu}\nabla_{\nu}\phi = \phi_{\mu\nu}$, $f_X = \partial f/\partial X$ and $f_\phi = \partial f/\partial \phi$, where $f$ is any function of $\phi$, $X$. 

The $\tilde{\mathcal{L}}_1$ term is called the tadpole term. It is not going to be considered any further, since it usually represents the standard cosmological constant and might lead to instabilities associated to vacuum energy \cite{TadpoleNeglect}.

\subsubsection{Generalized Galileon theory - Horndeski}
\label{SecGenGal}
In Sec.\ref{SecCovGal} we have seen how one obtains a covariant scalar-tensor theory possessing second-order field equations, starting from the Galileon theory. Similarly, one can start from the most general theory of scalar-field on a Minkowski background, which, unlike the Galileon theory, need not possess the Galilean shift symmetry, but should possess second-order field equations. Covariantizing this most general scalar-field theory in a similar way as in Sec.\ref{SecCovGal}, by adding appropriate counter terms, one obtains the generalized Galileon theory. In 4D, the corresponding Lagrangian is given by
\begin{align}
\mathcal{L}_{GG} &= G_2(\phi,X) - G_3(\phi,X)\Box\phi + G_4(\phi,X)R + G_{4X}\big[(\Box\phi)^2 - \phi^{\mu\nu}\tred{\phi_{\mu\nu}}\big] \nonumber \\
&+ G_5(\phi,X)G^{\mu\nu}\phi_{\mu\nu} - \frac{G_{5X}}{6}\big[(\Box\phi)^3 - 3\Box\phi \phi^{\mu\nu}\phi_{\mu\nu} + 2\phi_{\mu\nu}\phi^{\nu\lambda}\phi^{\mu}_{\lambda}\big]~,
\label{HorndeskiLag}
\end{align}
where $G_2,...,G_5$ are arbitrary functions of $\phi$ and $X$. This generalized Galileon theory was shown in \cite{genGalHorndeski} to be equivalent to Horndeski theory, which is the most general scalar-tensor theory possessing second-order field equations in 4D. Therefore, from here onwards, Eq.\eqref{HorndeskiLag} will be regarded as the Lagrangian for the Horndeski theories.

\subsubsection{Beyond Horndeski - DHOST theory}

In the previous Sec.\ref{SecGenGal}, we considered the most general scalar-tensor theory with second-order field equations in 4D. However, although having second-order field equations is a sufficient condition for a theory to be ghost-free, it is not a necessary one. Therefore, Horndeksi theories, having second-order field equations, are not the most general ghost-free theories. 

\begin{figure}[h]
	\centering
	\includegraphics[width=0.8\linewidth]{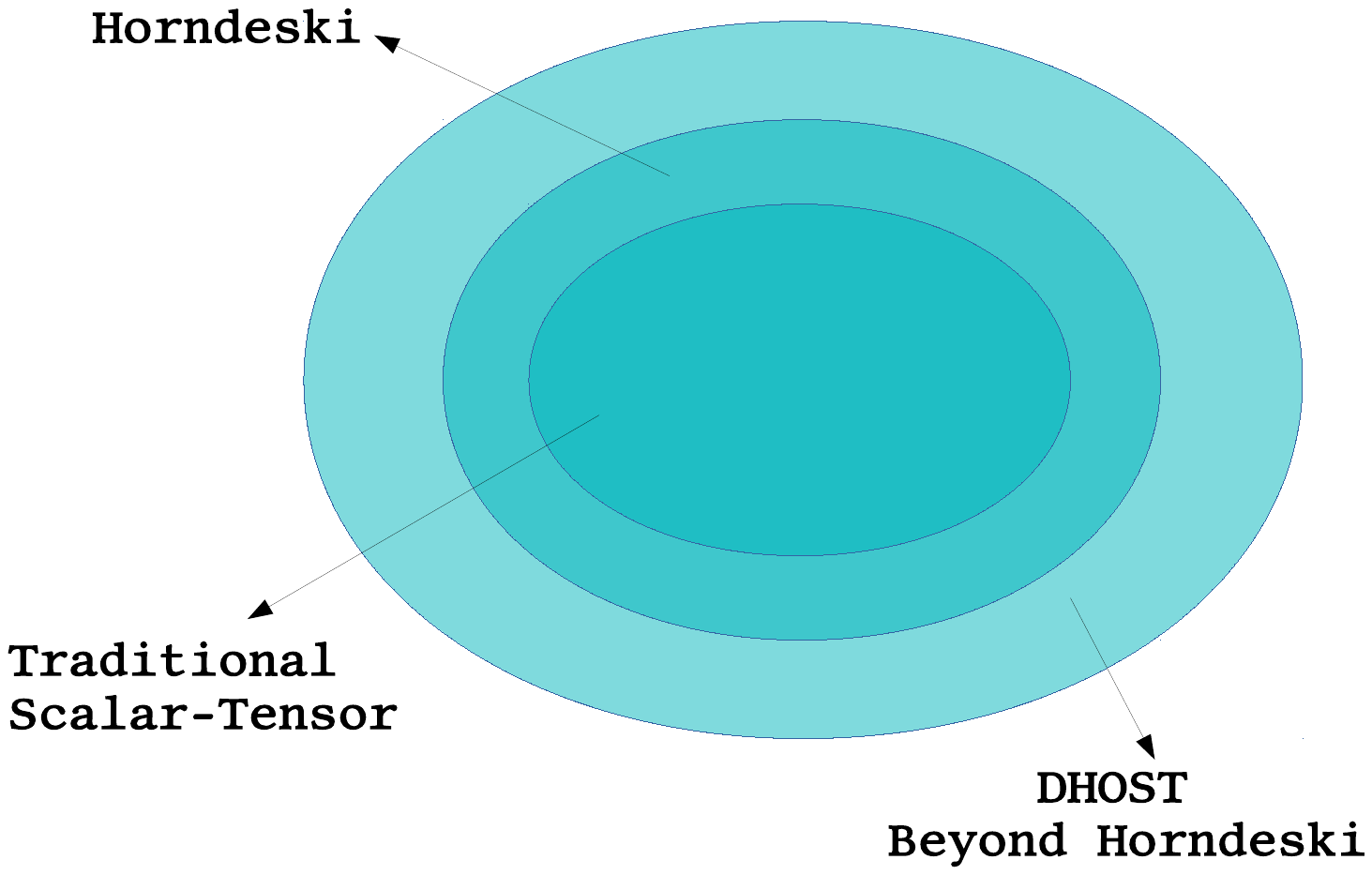}
	\caption{The traditional STTs are the subclass of Horndeski theories, which are further a subclass of DHOST theories beyond Horndeski. This figure is inspired by \protect\cite{DHOST_review2}.}
	\label{fig_STT}
\end{figure}

Recall that in classical mechanics, a theory can have higher-order (order higher than second) field equations and yet be free from Ostrogradsky-instabilities, provided the kinetic matrix constructed from the highest derivative terms in the field equations is degenerate. Such theories are called degenerate higher-order theories \cite{EvadingOstrogradsky}. Similarly, in the context of scalar-tensor theories, the most general ghost-free theories beyond Horndeski, are called degenerate higher-order scalar-tensor (DHOST) theories (see \cite{DHOST_review2} for a review). Such DHOST theories were formulated and classified up to quadratic order (in second-order derivatives of the scalar field), called quadratic DHOST theories \cite{DHOSTclassification1,DHOSTclassification2,DHOSTclassification3} and later up to cubic order in \cite{DHOSTcubic}, called cubic DHOST theories. Here we will be considering quadratic DHOST theories only, since the cubic DHOST theories are ruled out after GW170817. The Lagrangian for the quadratic DHOST theories is given by
\begin{equation}
\mathcal{L} = f(\phi,X)R + \sum_{i=1}^{5}A_{i}(\phi,X)L_i~,
\label{quadraticDHOST}
\end{equation}
where
\begin{align}
&L_1 = \phi_{\mu\nu}\phi^{\mu\nu}~,~~L_2 = (\Box\phi)^2~,~~ L_3 = \Box\phi\phi^{\mu}\phi^{\nu}\phi_{\mu\nu}~, \nonumber\\
&L_4 = \phi^{\mu}\phi_{\mu\alpha}\phi^{\alpha\nu}\phi_{\nu}~,~~L_5 = (\phi^{\mu}\phi^{\nu}\phi_{\mu\nu})^2~.
\end{align}
The above Lagrangian includes all the possible terms that are quadratic in the second derivatives of $\phi$. Upon comparing the above Eq.\eqref{quadraticDHOST} with the Horndeski Lagrangian Eq.\eqref{HorndeskiLag}, one finds that the quadratic DHOST Lagrangian is an extension to the Horndeski $G_4$ Lagrangian. One can include the $G_2(\phi,X)$, $- G_3(\phi,X)\Box\phi$ terms as well, since the degeneracy condition associated to the STT Lagrangian starting from which such DHOST theories are constructed, does not involve these two terms \cite{DHOST_review2,EvadingOstrogradsky}. An equivalent extension of the Horndeski $G_5$ Lagrangian results in the cubic DHOST theories, which will not be considered here, as stated earlier. Quadratic DHOST theories can be further classified into different subclasses, out of which only one particular class (we call this class Ia), which are related to the Horndeski through a disformal transformation $g_{\mu\nu}\rightarrow\tilde{g}_{\mu\nu} = C(\phi,X)g_{\mu\nu} + D(\phi,X)\phi_{\mu}\phi_{\nu}$, are deemed as viable \cite{DHOST_review2,HorndeskiBeyondReview}. This class is characterized by
\begin{align}
A_2 &= -A_1~,\\
A_4 &= \frac{1}{2(f+2XA_1)^2}[8XA_1^3 + (3f + 16Xf_X)A_1^2 - X^2fA_3^2 + 2X(4Xf_X - 3f)A_1A_3 \nonumber \\
&+ 2f_X(3f + 4Xf_X)A_1 + 2f(Xf_X - f)A_3 + 3ff_X^2]~,\\
A_5 &= -\frac{(f_X + A_1 + XA_3)(2fA_3 - f_XA_1 - A_1^2 + 3XA_1A_3)}{2(f+2XA_1)^2}~,
\end{align}  
with $f+2XA_1 \neq 0$. The terms $f$, $A_1$ and $A_3$ are independent functions of the theory in addition to the lower order terms $G_2$ and $G_3$. The following schematic diagram Fig.\ref{fig_STT} represents the broad classifications of STTs and their generalizations, where traditional STTs refer to the theories whose Lagrangians depend at most on first order derivatives of the scalar field (e.g., Brans–Dicke theory \cite{BransDicke}).

\subsubsection{Status of DHOST after GW170817}
\label{SecSurvivingDHOST}
The detection of gravitational waves GW170817 nearly at the same time with the $\gamma$-ray burst GRB 170817A from the binary neutron star merger constrains $c_{GW} = c$ to an accuracy of $\sim 10^{-15}$, where $c_{GW}$ refers to the speed of gravitational wave with $c=1$ being the speed of light. Therefore all the viable STTs must have to satisfy this constraint. The speed of gravitational waves in quadratic DHOST theories is given by
\begin{equation}
c_{GW}^2 = \frac{f}{f+2XA_1}~,
\label{cGWconstraint}
\end{equation}
which implies $A_1=0$, in order to satisfy the constraint imposed by GW170817. As far as the cubic DHOST theory is concerned, all cubic order (in second derivatives of $\phi$) terms lead to $c_{GW}\neq1$. Therefore the only viable DHOST theories after GW170817 constitutes the specific subclass of quadratic DHOST theories satisfying Eq.\eqref{cGWconstraint}
\begin{align}\label{survingDHOSTlag}
&A_2 = -A_1 = 0~,\\
&A_4 = \frac{1}{2f}[-X^2A_3^2 + 2(Xf_X - f)A_3 + 3\tred{f_X^2}]~, \label{survingDHOSTlag2}\\
&A_5 = -\frac{A_3(f_X + XA_3)}{f}~.
\label{parametersDHOST}	
\end{align}
Hence after GW170817, we are left with only two free functions $f$ and $A_3$, besides the lower order $G_2$ and $G_3$ terms. In this work we will be reviewing the implications of this class of viable DHOST theories post GW170817 event, on stellar and substellar objects. From this point onwards we will refer to such theories as {\it surviving}-DHOST theories.

\subsection{Vainshtein mechanism in surviving-DHOST theories}
\label{SecVainshtein}
In this section we will first review the basic principle of the Vainshtein mechanism, where nonlinear interactions in the Lagragian play an important role in screening the scalar degree of freedom on smaller scales, while allowing for modifications at larger scales. We then review the partial breaking of the Vainshtein mechanism in the surviving-DHOST theories. We will closely follow \cite{HorndeskiBeyondReview}.

\subsubsection{Principle of Vainshtein mechanism - nonlinearities}
\label{SecVainshteinNonlinear}
Let us consider a spherically symmetric non\tred{-}relativistic matter density, which perturbs a Minkowski background with a constant scalar field $\phi_0$. The matter density can be anything starting from a star to a cluster of galaxies, but whatever it may represent, we are interested in investigating how the scalar degrees of freedom in a STT mediate gravity around such matter densities. For that we consider a traditional STT Lagrangian (which is a subclass of DHOST theories, see Fig.(\ref{fig_STT})) containing only a non-minimally coupled scalar field term and a kinetic term. The corresponding action is
\begin{equation}
S = \int d^4x\sqrt{-g}[f(\phi)R + X] + S_m[g_{\mu,\nu}\psi_m]~.
\label{VainshteinAction}
\end{equation}
The matter fields (collectively represented by $\psi_m$) are minimally coupled to the metric $g_{\mu\nu}$, thus satisfying Einstein's Equivalence Principle.

The perturbed metric and the scalar field are given by
\begin{equation}
g_{\mu\nu} = \eta_{\mu\nu} + M_{Pl}^{-1}h_{\mu\nu}(t,\vec{x})~,~~ \phi = \phi_0 + \tred{\mathcal{\pi}(t,\vec{x})}~,
\end{equation}
where $M_{Pl} = 1/\sqrt{8\pi G_{N}}$ is the Planck mass, with $G_{N}$ being the Newton's constant.

Expanding the Lagrangian of Eq.\eqref{VainshteinAction} to the second order in both the metric and scalar field perturbations, and redefining the metric perturbation $h_{\mu\nu} = \tilde{h}_{\mu\nu} - \tred{2\bar{\xi}\mathcal{\pi}\eta_{\mu\nu}}$, one obtains the effective Lagrangian in the weak-field limit
\begin{equation}
\mathcal{L}_{eff} = -\frac{1}{4}\tilde{h}^{\mu\nu}\mathcal{E}_{\mu\nu}^{\alpha,\beta}\tilde{h}_{\alpha\beta} - \frac{(1+6\tred{\bar{\xi}^2})}{2}\partial_{\mu}\tred{\mathcal{\pi}}\partial^{\mu}\tred{\mathcal{\pi}} + \frac{1}{2M_{Pl}}\tilde{h}^{\mu\nu}T_{\mu\nu} - \frac{\tred{\bar{\xi}}}{M_{Pl}}\tred{\mathcal{\pi}}T~,
\label{EffectiveLagLinearVainshtein}
\end{equation}
with $T_{\mu\nu}$ and $T$ being the energy momentum tensor of the matter density and its trace, respectively. $\mathcal{E}_{\mu\nu}^{\alpha,\beta}\tilde{h}_{\alpha\beta}$ is the linearized Einstein tensor and
\begin{equation}
\tred{\bar{\xi}} = M_{Pl}^{-1}df/d\phi|_{\phi=\phi_0}~.
\end{equation}
Integrating the field equations in both the metric and scalar field, derived from the above Lagrangian, one obtains the metric potentials outside the matter distribution
\begin{equation}
\Phi = -\frac{G_{N}M}{r}~,~~\Psi = \gamma\Phi~,
\label{LinearVainshteinPotentials}
\end{equation}
where $h_{00} = -2\Phi$ and $h_{ij} = -2\Psi\delta_{ij}$. The Newton's constant $G_{N}$ and $\gamma$ are functions of the scalar field appearing in the theory, and $M$ is the total mass of the matter distribution.

From Eq.\eqref{LinearVainshteinPotentials}, we observe that $\Phi\neq \Psi$, and therefore, the metric potentials outside the matter do not conform to that of the metric for a star or planet in the weak-field limit of GR \cite{CarrollGR},
\begin{equation}
ds^2 = -(1+2\Phi)dt^2 + (1-2\Phi)(dx^2 + dy^2 + dz^2)~,
\end{equation}
which could successfully predict the planetary orbits and trajectory of light rays in the gravitational field of the Sun \cite{Blau}. Therefore, the effective Lagrangian of Eq.\eqref{EffectiveLagLinearVainshtein} contradicts the solar-system experiments. However, if one introduces a nonlinear cubic interaction term 
\begin{equation}
-\frac{1}{2\tred{\bar{\Lambda}}^3}(\partial\tred{\mathcal\pi})^2\Box \tred{\pi}~,
\label{cubicInteractionTerm}
\end{equation}
with $\tred{\bar{\Lambda}}$ \tred{$\sim (H_{0}^{2}M_{Pl})^{1/3}$} as an appropriate mass scale in Eq.\eqref{EffectiveLagLinearVainshtein}, and define a length scale called the Vainshtein radius
\begin{equation}
r_V = \Big(\frac{M}{8\pi M_{Pl}\tred{\bar{\Lambda}}^3}\Big)^{1/3}~,
\end{equation}
then one recovers weak-field limit of GR (i.e., $\Phi\simeq\Psi$) for $r\ll r_V$, while the modified potentials Eq.\eqref{LinearVainshteinPotentials} prevail for $r\gg r_V$. \tred{$H_{0}$ above is the present-day Hubble constant.} Considering the matter density to be a solar mass star, and using the present day values of the accelerating expansion of the universe, one obtains $r_V \sim 100 pc$, which is way larger than the radius of our solar system. Therefore, we see that the nonlinear interaction Eq.\eqref{cubicInteractionTerm} is crucial in screening the modifications mediated by the scalar field, in the solar system scale, while allowing such modifications at larger scales.

\subsubsection{Partial breaking of Vainshtein mechanism}
\label{Secpartialbreaking}
In the previous subsection, we have seen how nonlinear interaction terms in the effective Lagrangian are important in screening scalar degree of freedom at small scales. Such terms are by default present in the surviving-DHOST theories, and therefore the Vainshtein screening prevails. However, inside the matter density, where the mass is not a constant, the standard gravitational potential gets modified, therefore leading to the breaking of the Vainshtein mechanism.

Unlike in the previous subsection where we studied the metric and scalar field perturbations around a Minkowski background with a constant scalar field, here one considers the Vainshtein mechanism from a more realistic cosmological perspective. We take Newtonian perturbations around an Friedman-Robertson-Walker (FRW) space-time,
\begin{equation}
ds^2=-[(1+2\Phi(t,x^i))]dt^2 + a(t)^2[(1\tred{-}2\Psi(t,x^i))]\delta_{ij}dx^idx^j~,
\label{FRWnewtonMetric}
\end{equation}
with a perturbed scalar field $\phi(r,t) = \phi_0(t) + \pi(r,t)$, where $\phi_0(t)$ is the time dependent background scalar field. The Lagrangian corresponding to the surviving-DHOST theories Sec.\ref{SecSurvivingDHOST}, is then chosen \cite{CrisostomiKoyama}.

Similar to the above formalism in Sec.\ref{SecVainshteinNonlinear}, one derives the effective Lagrangian, where the nonlinear terms important for Vainshtein screening are retained, while the terms involving time-derivatives of the field and metric perturbations are ignored under the quasi-static approximation \cite{KoyamaSakstein,HorndeskiBeyondReview,CrisostomiKoyama}. By integrating the field equations derived from the resulting effective Lagrangian, the following is obtained for $r\ll r_V$ \cite{HorndeskiBeyondReview,CrisostomiKoyama} :
\begin{align}\label{ModPoisson1}
\frac{d\Phi}{dr} &= G_N \Big(\frac{M\tred{(r)}}{r^2} + \frac{\Upsilon_1 M\tred{(r)}''}{4}\Big)~,\\
\frac{d\Psi}{dr} &= G_N \Big(\frac{M\tred{(r)}}{r^2} - \frac{5\Upsilon_2 M\tred{(r)}'}{4r} + \Upsilon_3M\tred{(r)}''\Big)~,
\label{ModPoisson2}
\end{align}
with a prime $'$ denoting a derivative with respect to the radial coordinate $r$, and where
\begin{equation}
G_N = \frac{1}{8\pi} \big[2(f-Xf_X-3X^2A_3)\big]^{-1}~,
\end{equation}
\begin{equation}
\Upsilon_1 = -\frac{(f_X-XA_3)^2}{A_3f}~,~~\Upsilon_2=\frac{8Xf_X}{5f}~,~~\Upsilon_3=\frac{(f_X-XA_3)(f_X+XA_3)}{4A_3f}~.
\label{UpsilonExpression}
\end{equation}
\tred{The functions $X$, $f$, $f_X$, and $A_3$ are all evaluated at the background, i.e., $\phi=\phi_0(t)$ \cite{HorndeskiBeyondReview,CrisostomiKoyama}, and as a consequence, the parameters $\Upsilon_1,\Upsilon_2$, and $\Upsilon_3$ are spatially constant at a given time $t$.}

We therefore see from Eqs.\eqref{ModPoisson1} and \eqref{ModPoisson2} that the Vainshtein mechanism works outside the matter density, where $M\tred{(r)(=M)}$ is constant, but breaks inside the matter density where $M\tred{(r)}$ is not a constant. The parameters $\Upsilon_1,\Upsilon_2$ and $\Upsilon_3$ quantify the deviations from the standard gravity, where $\Upsilon_1$, specifically controls the modification of the Newtonian potential, and thus the corresponding pressure balance equation. Since the pressure balance equation in the Newtonian limit is an integral part of the analysis of stellar and substellar dynamics and evolution, this parameter can therefore be constrained through stellar observables in the low-energy limit.

For a special choice $A_3=-f_X/X$ \cite{HorndeskiBeyondReview} of the free parameters in the surviving-DHOST theories Eq.\eqref{parametersDHOST}, which represents the Gleyzes-Langlois-Piazza-Vernizzi (GLPV) family \cite{GLPV1,GLPV2}, the equations for the metric potentials Eq.\eqref{ModPoisson1}, \eqref{ModPoisson2} reduce to
\begin{align}\label{GLPVPoisson1}
\frac{d\Phi}{dr} &= G_N \Big(\frac{M\tred{(r)}}{r^2} + \frac{\Upsilon_1 M\tred{(r)}''}{4}\Big)~,\\
\frac{d\Psi}{dr} &= G_N \Big(\frac{M\tred{(r)}}{r^2} - \frac{5\Upsilon_2 M\tred{(r)}'}{4r}\Big)~.
\label{GLPVPoisson2}
\end{align}
Therefore in GLPV theories, only two parameters $\Upsilon_1$, and $\Upsilon_2$ quantify the deviations from standard Einstein gravity. The parameter $\Upsilon_2$ can be constrained through relativistic observations \cite{CrisostomiKoyama,SaksteinWilcox}.

\section{Stellar modeling with modified gravity and anisotropies}
\label{SecModAnisotropy}
In this section we will describe how stellar anisotropies modify the defining stellar structure equations in modified gravity theories while retaining the approximation of spherical symmetry. We will briefly state the generic numerical prescription to obtain the complete stellar model from such equations. Then we will show how such anisotropies, in general, constrain the modified gravity
parameter. Finally, we will discuss the results and constraints
obtained in considering such modified equations inside different classes of stellar
objects.

\subsection{Modified stellar equations and numerical recipe}
\label{SecMHEC}
Any stellar pressure anisotropy in {\it spherically symmetric} fluids modifies the corresponding isotropic pressure balance equation by an additive term, which denotes the measure of anisotropy \cite{BowersLiang}. In this subsection we derive the expression for the hydrostatic equilibrium condition for spherically symmetric fluids, in presence of stellar pressure anisotropy, within the framework of modified gravity theories \cite{Chowdhury1,Chowdhury2}. We then briefly describe the modeling of such an additive anisotropy term based on the source of the anisotropy. Finally we conclude this subsection by outlining the numerical prescription that needs to be followed in general, to obtain the complete stellar model.

\subsubsection{Modified hydrostatic equilibrium condition}

In the Newtonian limit, the stress energy tensor inside a spherically symmetric anisotropic stellar object is taken to be $T^{\mu}_{\nu}=diag(-\rho,P_r,P_\perp,P_\perp)$, where $P_r$ corresponds to the radial pressure and $P_\perp$ is the tangential one. Now, the covariant conservation of this energy momentum tensor, i.e., $D_{\mu}T^{\mu\nu}=0$ ($D_\mu$ being the covariant derivative) in the Newtonian gauge Eq.\eqref{FRWnewtonMetric} in a static situation gives \cite{Chowdhury2}
\begin{equation}
\frac{dP_{r}}{dr} = -\rho \frac{d\Phi}{dr} +\frac{2}{r}\left(P_{\perp}-P_{r}\right)\left(1-r\frac{d\Psi}{dr}\right)~.
\label{HEC}
\end{equation}
Although the terms dependent on $\Psi$ affect the pressure balance equation for anisotropic astrophysical objects, it was shown in \cite{Chowdhury1} that such terms can be safely ignored for the white dwarfs, and brown dwarfs in the context of GLPV theories; similar arguments may apply for intermediate mass stars as well. However, such terms might be relevant for galaxies or galaxy clusters. Therefore, using Eq.\eqref{GLPVPoisson1} for GLPV theories, one obtains the modified hydrostatic equilibrium condition (MHEC), inside anisotropic stellar and substellar objects
\begin{equation}
\frac{dP_r}{dr} = - \frac{G_NM\tred{(r)}\rho}{r^2} - \Upsilon\left(\frac{G_N\rho}{4}\right)\frac{d^2M\tred{(r)}}{dr^2} + \Delta(r)~,
\label{MHEC}
\end{equation}
with $\Upsilon=\Upsilon_1$ from here on for ease of notation, and $\Delta(r)= 2(P_{\perp}-P_{r})/r$ being the measure of anisotropy.

\subsubsection{Modeling stellar pressure anisotropy}

To proceed further, one needs to choose a specific form of $\Delta(r)$. Out of the several reasons that lead to such stellar pressure anisotropy \cite{Herrera1997}, the important ones that will be relevant in the context of stellar and substellar physics are the following: (1) stellar fluid composed of two (or more) isotropic fluids exhibit such pressure anisotropy $(P_\perp-P_r)\sim {\rm function~of} ~(\rho_1,\rho_2,P_1,P_2,U^{\mu},W^{\mu})$, where the subscripts $-1,2$, identifies the quantities to the individual fluids with $U^{\mu},W^{\mu}$ being the four velocities of the two fluids; (2) slowly rotating fluid, can be modeled by such pressure anisotropy term $(P_\perp-P_r)\sim \rho\Omega^2r^2$, due to centrifugal forces, to the first order approximation \cite{Kippenhahn,Herrera1997}, where $\Omega$ denotes angular rotation speed; (3) time-independent stellar magnetic fields $B$, of small magnitudes, can also be approximated by such pressure anisotropy term $(P_\perp-P_r)\sim B^2$ \cite{Ferrer2010}. 

Although the anisotropies due to two fluid approximation are useful in modeling Neutron stars \cite{Letelier,Bayin}, in the context of the stellar models that will be considered here, rotation and magnetic fields will be crucial. Since stellar rotation breaks the spherical symmetry and a simplistic anisotropy term in the pressure balance equation, retaining the spherical symmetry might not be sufficient; such a case will be dealt later in Sec.\ref{SecRotation}. However, when the specific form of $\Delta(r)$ is chosen according to the model $(P_\perp-P_r)=\beta(r)P_r(r)$, motivated by \cite{HeintzmannHillebrandt}, the resulting anisotropy can approximate the situation where the stellar pressure anisotropy originates at least partly from magnetic fields (see \cite{Chowdhury1}). In the above expression, $\beta(r)$ represents the dimensionless
measure of the strength of anisotropy. Its form is given as $\tred{\beta(r)=\tau (r/r_0)^{2}}$ with
$\tau$ being referred to as the anisotropy parameter and $r_0$ is an appropriate length
scale, whose expression depends largely on whether the stellar matter is modeled
through polytropic equation of state or non-polytropic one \cite{Chowdhury4}. It is to be noted that the above modified hydrostatic equilibrium condition Eq.\eqref{MHEC} is valid as long
as the approximation of spherical symmetry is well justified; for that the value of
the anisotropy parameter will be kept low throughout \cite{Chowdhury1,Chowdhury4}

\subsubsection{Other stellar structure equations}
The modified pressure balance equation mentioned above, along with the mass conservation equation $dM/dr=4\pi r^2\rho$, the radiative transfer equation 
\begin{equation}
\frac{dT(r)}{dr}=-\frac{3}{4a}\frac{\kappa}{T(r)^3}\frac{\rho(r) L(r)}{4\pi r^2}~,
\label{RTE}
\end{equation}
 and the energy transport condition
\begin{equation}
\frac{dL(r)}{dr}=4\pi r^{2}\rho(r)\epsilon(r)~,
\label{ETC}
\end{equation}
forms the crucial ingredients for studying anisotropic stellar physics within the framework of modified gravity theories \cite{KoyamaSakstein,Chowdhury2,Chowdhury4}. In the above Eqs.\eqref{RTE},\eqref{ETC}, $L(r)$ is the stellar luminosity, $\kappa$ denotes the opacity, $\epsilon(r)$ is the energy released per unit mass per unit time, with $a$ being the radiation-density constant. Besides these stellar structure equations, one needs the following constitutive relations \cite{CarrollOstlie},
\begin{align}
P &= P(\rho,T,{\rm composition})~,\label{EOS}\\
\kappa &= \kappa(\rho,T,{\rm composition})~,\\
\epsilon &= \epsilon(\rho,T,{\rm composition})~,
\end{align}
where the pressure, opacity and energy generation rates are expressed in terms of the fundamental properties of the stellar material. From here onwards, we will refer to Eq.\eqref{EOS} as the equation of state (EOS), governing the matter inside stellar or substellar objects.

\subsubsection{Generic numerical recipe}
To obtain the complete structure for stellar objects with homogeneous composition, one integrates the above set of stellar structure equations with specific boundary conditions at $r=0$ (center) and at $r=R$ (stellar surface):
\begin{equation}
M(0)=0~,~T(0)=T_{c}~,~P(0)=P_{c}~;~~ M(R)=M~,~T(R)=0=P(R)~,
\label{BC}
\end{equation}
where $T_c$ and $P_c$ correspond to central temperature and pressure respectively. However, for stellar objects with distinctive core envelope structure, there exists two sets of stellar structure equations, one for the core and the other for the envelope. The equations on the core side needs to be integrated outwards given the boundary condition at the center, while those on the envelope side needs to be integrated inwards corresponding to the boundary condition at the stellar surface. Now, in order to ensure the continuity of stellar structure variables like mass, pressure and radius, at the core envelope junction, one needs to employ the following fitting condition for homology invariants $U$ and $V$ defined as
\begin{equation}
U=\frac{d \ln M(r)}{d \ln r}~,~~V=-\frac{d \ln P(r)}{d \ln r}~.
\label{UV}
\end{equation}
They satisfy the conditions
\begin{equation}
U_{fe}=\frac{1}{\alpha} U_{fc}~,~~V_{fe}=\frac{1}{\alpha} V_{fc}~.
\label{UVfit}
\end{equation}
at the fitting point, which is the core-envelope junction (see \cite{Kippenhahn}). The subscript "fc" denotes the fitting point, approached from the core side, with the subscript "fe" denoting the fitting point when approached from the envelope side.

This apparently simple minded algorithm entails a plethora of subtleties in reality, which needs to be taken care of carefully. For example, some variable re-definitions become important in reducing the number of degrees of freedom of the stellar model, as well as to carry out numerical integration near points where the stellar equations become singular \cite{Chowdhury2,Chowdhury4}. For a more in-depth understanding of the numerical recipe for solving stellar structure equations in different stellar models, see the excellent book \cite{Schwarzschild}.

\subsection{Generic constrain on $\Upsilon$ due to anisotropy}
\label{SecGenericConstraintUpsTau}
In this section we will show how stellar pressure anisotropy, in general, constrains the modified gravity parameter, within the approximation of spherical symmetry.

\begin{figure}[h]
	\hspace{0.3cm}
	\begin{minipage}[t]{0.40\linewidth}
		\centering
		\centerline{\includegraphics[scale=0.24]{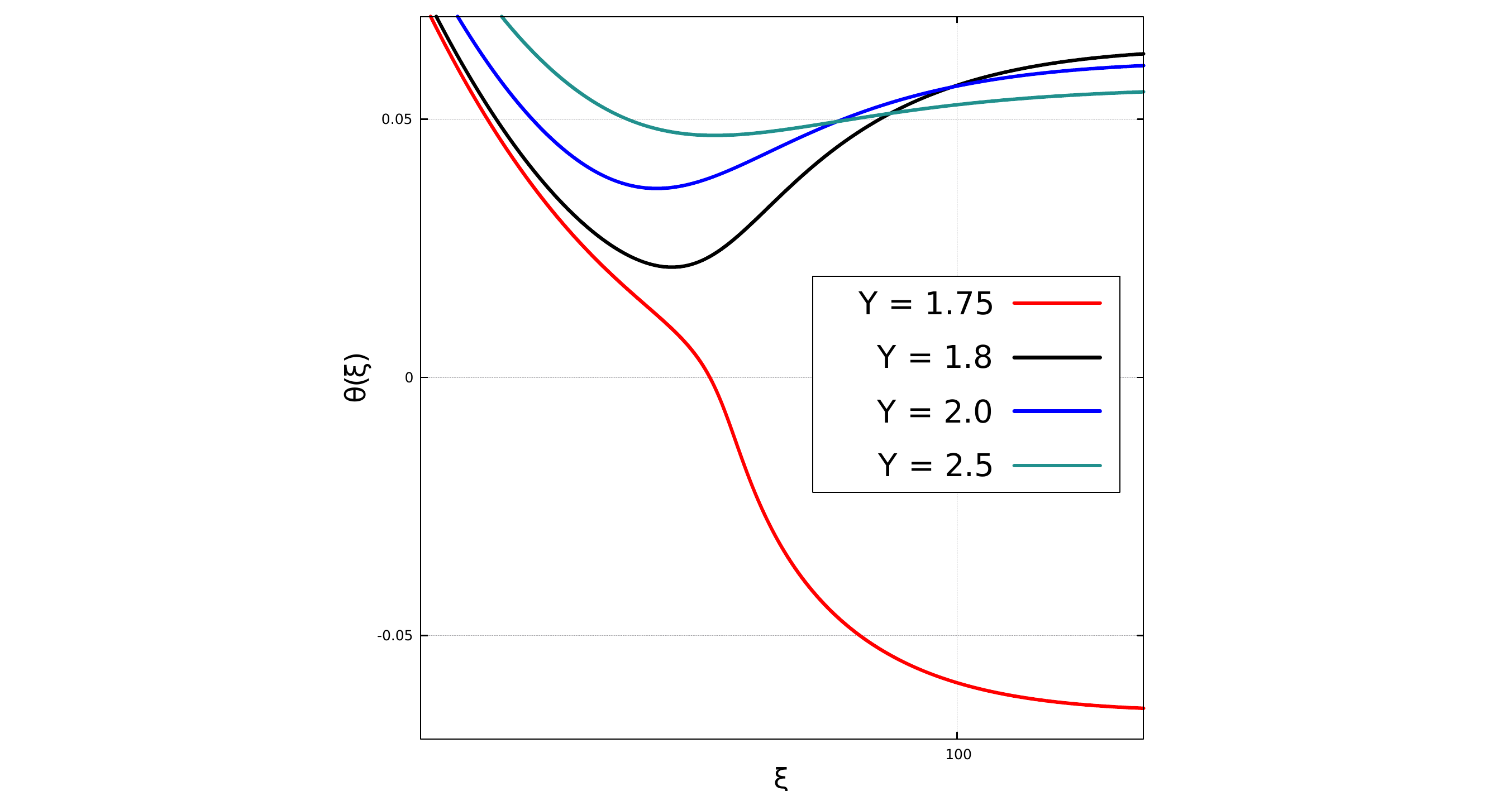}}
		\caption{$\theta$ vs $\xi$ plots for $ n= 3$ and $\tau = 0.01$ \protect\cite{Chowdhury1}.}
		\label{Fig_turning}
	\end{minipage}
	\hspace{0.9cm}
	\begin{minipage}[t]{0.40\linewidth}
		\centering
		\centerline{\includegraphics[scale=0.24]{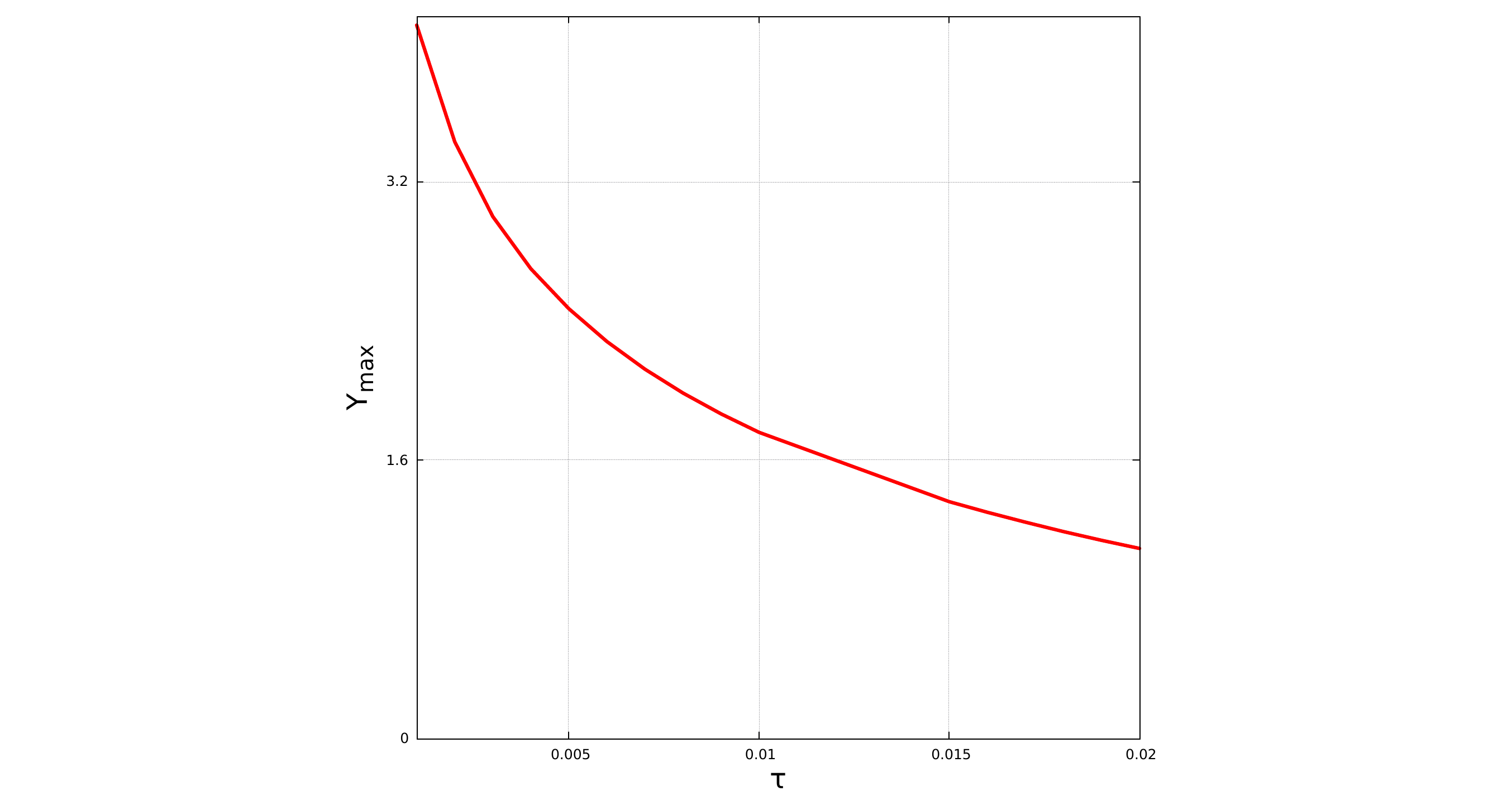}}
		\caption{$\Upsilon_{\rm max}$ vs $\tau$ plots for $n = 3$ \protect\cite{Chowdhury1}.}
		\label{Fig_upsmax}
	\end{minipage}
\end{figure}

The polytropic EOS for the anisotropic stellar or substellar matter \cite{Shapiro1983} is given as
\begin{equation}
P_r = K \rho^{\frac{n+1}{n}}~,
\label{poly}
\end{equation}
with $n$ being the polytropic index and $K$ being the polytropic constant; their values and expressions depend upon the class of
stellar or substellar objects being considered.
In terms of the dimensionless variables $\theta$ and $\xi$ defined as
\begin{equation}
\rho=\rho_c \theta^n~,~~~ r=r_c\xi~,~~~{\rm with}~~~~ r_c^2=\frac{K(n+1)\rho_c^{(\frac{1}{n}-1)}}{4\pi G_N}~,
\label{nondimensional0}
\end{equation}
with $\rho_c$ being the central density, and with the form of $\Delta(r)$ mentioned in Sec.\ref{SecMHEC}, the MHEC Eq.\eqref{MHEC} leads to the modified Lane-Emden equation (MLEE) \cite{Chowdhury1,Chowdhury4}
\begin{equation}
\frac{1}{\xi^2}\frac{d}{d\xi}\left[\xi^2\frac{d\theta}{d\xi} - \frac{2}{n+1}\tau\xi^3\theta + \frac{\Upsilon}{4}\left(2\xi^3\theta^n + n\xi^4
\theta^{n-1}\frac{d\theta}{d\xi} \right)\right] + \theta^n=0~.
\label{MLEE}
\end{equation}
The condition of local maxima of the stellar pressure at the center yields \cite{Chowdhury1}
\begin{equation}
\Upsilon > -\frac{2}{3} + \frac{4\tau}{n+1}~,
\label{upscon1}
\end{equation}
while the condition for the existence of local minima in the stellar pressure, away from the center but well within the stellar radius, yields \cite{Chowdhury1}
\begin{equation}
\Upsilon \lesssim -\frac{2}{3} + \frac{4\tau}{n+1} \theta_T^{1-n}~,
\label{upscon2}
\end{equation}  
where $\theta_T$ refers to the value of $\theta$ at its local minima. While the pressure is indeed maximum at the stellar center for any viable model, the existence of local pressure belts is quite unphysical. However it was shown in \cite{Chowdhury1} that for a given $n\geq1$, such pressure belts are observed for values of $\Upsilon$ higher than a certain maximum for any value of $\tau>0$, see Fig.\ref{Fig_turning}. Therefore, for every given positive $\tau$, one obtains an upper bound on $\Upsilon$ due to the stellar pressure anisotropy. We will denote this upper bound as $\Upsilon_{\rm max}$, see Fig.\ref{Fig_upsmax}.

Although such local minima are theoretically possible for any value of $n\geq1$, it is seen that for lower values of $n$, such minima exists only for $\Upsilon$ and $\tau$ values which makes the stellar radius unrealistically large \cite{Chowdhury1}.

\subsection{Results and constraints from different stellar and substellar objects}

Now we discuss the main results and constraints obtained by considering the modified stellar equations discussed in Sec.\ref{SecMHEC} inside different classes of stellar and substellar objects. We will start from VLM objects in the pre-main sequence phase, and then move along the evolutionary track up to the white dwarf scenario, while briefly mentioning the status and challenges pertaining to the evolutionary phases from SGB to AGB.

\subsubsection{VLM objects}
\label{SecVLM}
VLM objects are fully convective, and are therefore modeled as polytropes $P=K\rho^{5/3}$, with $n=3/2$ and the polytropic constant $K$ depending upon the degeneracy parameter $\eta$ as follows \cite{BurrowsLiebert}
\begin{equation}
K=\frac{(3\pi^2)^{2/3}\hbar^2}{5m_{e}m_{p}^{5/3}\mu_{el}^{5/3}}\left(1+\frac{\tilde{\alpha}}{\eta}\right)~,~~ \eta=\frac{(3\pi^2)^{2/3}\hbar^2}{2m_{e}m_{p}^{2/3}k_{B}\mu_{el}^{2/3}}\frac{\rho^{2/3}}{T}~.
\label{VLMPolytrope}
\end{equation}
Here, $\eta$ is defined as the ratio of Fermi energy to thermal energy. In Eq.\eqref{VLMPolytrope}, $m_e$ and $m_p$ are the masses of the electron and proton, respectively, and $k_B$ is Boltzmann's constant. Here, we use $\tilde{\alpha} = 5\mu_{el}/(2\mu)$, 
where $\mu$ is the mean molecular weight of helium and the partially ionized hydrogen
mixture in the interior, and $\mu_{el}$ is the number of baryons per electron. 

The energy generation rate of a truncated pp chain reaction inside VLM objects with characteristic central temperature $\sim 10^6$K and density $\sim 10^3$${\rm g/cm^3}$, can be modeled as a power law in $T$, and $\rho$ \cite{BurrowsLiebert} as
\begin{equation}
\label{energygen}
\epsilon=\epsilon_c\Big(\frac{T}{T_c}\Big)^s\Big(\frac{\rho}{\rho_c}\Big)^{u-1}~,~~
\epsilon_c=\epsilon_0 T_c^s\rho_c^{u-1}~ergs~g^{-1}s^{-1}~,
\end{equation}
with $s \simeq 6.31$, $u\approx2.28$ and $\epsilon_0=1.66\times10^{-46}$. With this information, one can calculate the hydrogen burning luminosity $L_{{\rm HB}}$.

Also, the photospheric temperature $T_e$, and density $\rho_e$ are related as 
\begin{equation}
\label{photoTrho}
\frac{T_e}{{\rm K}}=\frac{1.8\times10^6}{\eta^{1.54}}\Big(\frac{\rho_e}{{\rm g/cm^3}}\Big)^{0.42}~,
\end{equation}
which is obtained from second order plasma phase transition (PPT) near the photosphere, from metallic hydrogen to molecular hydrogen \cite{BurrowsLiebert}. This photospheric relation helps in calculating surface luminosity $L_{\rm s}$.

\begin{figure}[h!]
	\centering
	\centerline{\includegraphics[scale=0.24]{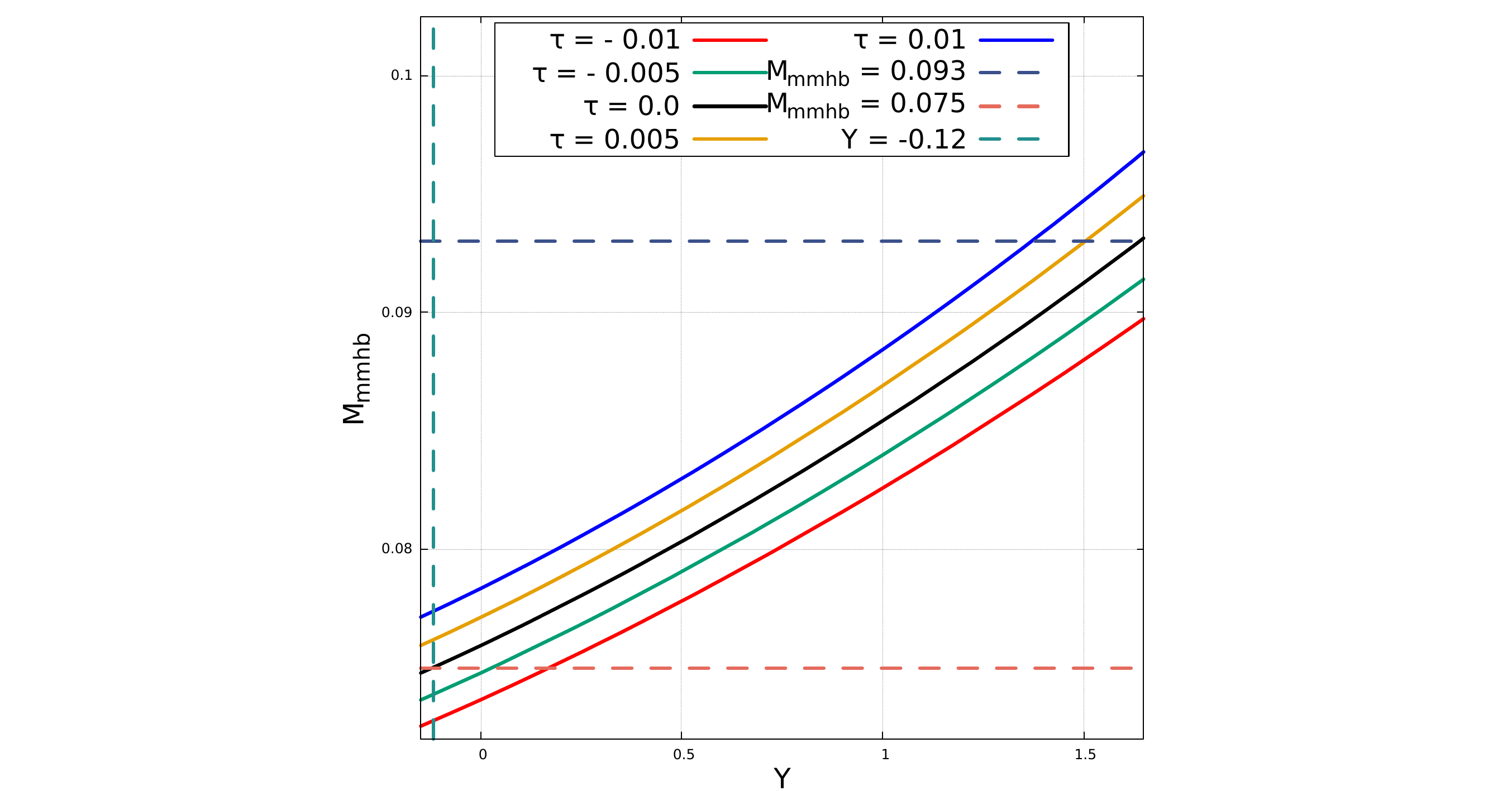}}
	\caption{Minimum mass of hydrogen burning (in units of $M_{\odot}$) in VLM objects as a function of $\Upsilon$ for different $\tau$, with $n = 3/2$ \protect\cite{Chowdhury1}.}
	\label{FigMmmhbAnisoMod}
\end{figure}

Now, for different values of the modified gravity parameter $\Upsilon$ and anisotropy parameter $\tau$, one obtains the lowest possible mass value, for which the stable hydrogen burning condition $L_{\rm HB}=L_{\rm s}$, is attained. This value corresponds to the $M_{\rm mmhb}$. It was first observed in \cite{Chowdhury1} that higher values of $\tau$ as well as $\Upsilon$, correspond to higher $M_{\rm mmhb}$ values, see Fig.\ref{FigMmmhbAnisoMod}. The reason is that higher values of anisotropy as well as modified gravity parameter leads to weakening of gravitational strength inside VLM objects. This makes the star achieve hydrostatic equilibrium at a lower temperature and density, and therefore the hydrogen burning luminosity is lowered. Hence more mass is needed to attain a stable hydrogen burning condition.  

The lowest mass main sequence stars (called M-dwarfs because of its spectral classification) are observed to be $\sim 0.093M_{\odot}$ \cite{Segransan2000}. Therefore, $M_{\rm mmhb}$ values higher than this needs to be ruled out. For the isotropic case, an upper bound of $\Upsilon = 1.6$ (see Fig.\ref{FigMmmhbAnisoMod}) is obtained \cite{SaksteinPRL, Chowdhury1}. It is generally believed that $M_{\rm mmhb}$ should range between $0.075M_{\odot}$ and $0.08M_{\odot}$. As a result, in the isotropic scenario, values of $M_{\rm mmhb}$ below $0.075M_{\odot}$ are ruled out, thereby yielding a lower bound of $\Upsilon = -0.12$ (shown by vertical dashed line in Fig.\ref{FigMmmhbAnisoMod}). In isotropic scenarios, other works related to the pre-main sequence evolution of VLM objects and low mass stars, within the framework of STTs, have been reported in \cite{WojnarVLMearlyEvolutionSTT, WojnarBDcoolingSTT}, while \cite{WojnarEarlyEvolutionPalatini,WojnarBDcoolingPalatini,WojnarPreMainSeqEiBI,WojnarLithium,WojnarMmmhbPalatini} study within the framework of other modified gravity theories like EiBI gravity and Palatini $f(R)$ gravity.

\subsubsection{Low mass stars (near solar mass) at main-sequence}

In isotropic scenarios, the main-sequence phase of near solar-mass stars have been studied in \cite{KoyamaSakstein} within the framework of beyond Horndeski (GLPV) theories. Using the Eddington standard model approximation, such main-sequence stars are modeled by $n=3$ polytropes,
\begin{equation}
P=K(\tilde{\beta})\rho^\frac{4}{3}~.
\end{equation}
The polytropic constant depends upon the ratio $\tilde{\beta}$ of the gas pressure to the total pressure, where the total pressure is due to both thermal and radiation pressures. Since higher values of $\Upsilon$ weakens the gravitational strength within the stellar object, hydrostatic equilibrium is maintained at lower central temperature and densities. This results in lowered energy production due to pp chain reaction at the stellar core, and thus the overall stellar luminosity decreases. The authors of \cite{KoyamaSakstein} have shown that the effects of modified gravity on stellar luminosity are much more prominent in near solar-mass main-sequence stars than those having higher masses. The reason being the stellar luminosity of stars supported by gas pressure is more sensitive to the strength of gravity, and near solar-mass stars are dominated by gas pressure. However, such stellar objects have not yet been analyzed in presence of anisotropies, and therefore the topic is relevant for future studies.

\subsubsection{Low mass stars (near solar-mass) at turnoff point}
\label{SectionpopII}
\begin{figure}[h]
	\centering
	\includegraphics[width=0.7\linewidth]{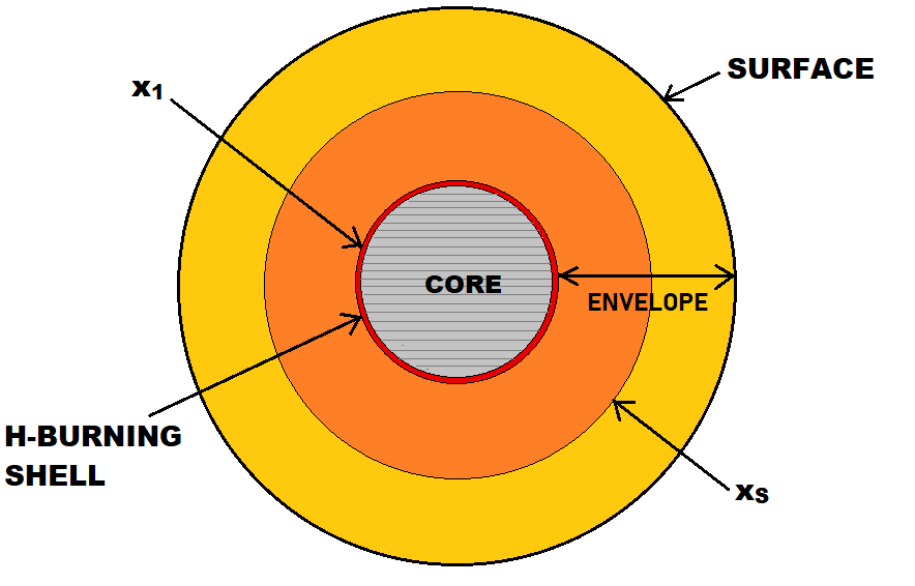}
	\caption{Schematic Diagram for the stellar model \protect\cite{Chowdhury2}.}
	\label{FigTurnoffSchematic}
\end{figure}

In this section, we discuss the work of \cite{Chowdhury2}, in which modified gravity was considered inside isotropic stars having a distinct core-envelope structure, with a thin hydrogen burning shell in between. Such a stellar structure model is relevant for low mass stars at their turnoff point. Since the prime interest in this work was to decipher the role of modified gravity on the stellar variables, the metallicity was neglected, to put aside any degeneracy that may arise out of it. In this work, the phenomenological model of a metal-poor population II star at its turnoff point as given by \cite{HS} was considered, with the following standard assumptions \cite{Schwarzschild}:
\begin{enumerate}
	\item The radiation pressure and relativistic degeneracies are negligible.
	\item The isothermal helium core is partially degenerate.
	\item The hydrogen rich envelope is in radiative equilibrium.
	\item The core mass fraction is taken to be $0.10$.
	\item The thickness of the hydrogen-burning shell is assumed to be negligible.
	\item The envelope is assumed to consist of two parts having distinct opacities:
	\begin{enumerate}
		\item The inner part having higher temperature is governed by opacity due to free electron scattering.
		\item The outer part having lower temperature is governed by opacity due to free-free transitions.
	\end{enumerate}
	\item The temperature and pressure at the stellar surface are taken to be zero.
\end{enumerate}
The schematic representation of this particular stellar model is given in Fig.\ref{FigTurnoffSchematic}
where $x_1$ denotes the non-dimensional radial coordinate of the hydrogen-burning shell, intermediate to the core and envelope. In the envelope, $x_s$ denotes the non-dimensional radial coordinate, where the mechanism of opacity changes.

The pressure and density inside the partially degenerate isothermal core is given by \cite{Chowdhury2}
\begin{equation}
P=\frac{8\pi}{3h^{3}}(2m_ek_BT)^{3/2}k_BTF_{3/2}(\psi)~,~~
\rho=\frac{4\pi}{h^{3}}(2m_ek_BT)^{3/2}\mu_{c}m_pF_{1/2}(\psi)~,
\end{equation}
where $F_{\nu}(\psi)$ is the Fermi function defined as
\begin{equation}
F_{\nu}(\psi)=\int_{0}^{\infty} \frac{u^{\nu}}{e^{u-\psi}+1} du~,
\label{Fermifunction}
\end{equation}
with $\psi$ being the degeneracy function \cite{HS} (see also \cite{DeMarque}). Although the ion pressure inside the core is not considered in this particular model, including it however, will not result in significant differences in stellar observables \cite{Hayashi}.  

The pressure in the non-degenerate radiative envelope is given by the ideal gas law
\begin{equation}
P=\frac{k_B}{\mu_e m_p}\rho T~.
\label{idealgas}
\end{equation}

The above stellar model entails a large number of algebraic and coupled differential equations, which needs to be solved with appropriate boundary and junction conditions to obtain the complete stellar structure. This is done using homology invariants as elaborated in \cite{Chowdhury2}.
To further simplify the numerical computations, the less dominant source of energy in the hydrogen-burning shell, i.e., the pp chain reaction is neglected \cite{Chowdhury2}. 

From numerical analysis, it is found that such near solar-mass stars at their turnoff point, becomes larger (their radius $R$ increases), cooler (their effective temperature $T_{\rm eff}$ decreases), and dimmer (surface luminosity $L$ decreases), for higher values of the modified gravity parameter $\Upsilon$. However, the core radius decreases with an increase in the temperature and density at the core-envelope junction, on the envelope side. This seemingly counter-intuitive behaviour is well justified from the fact that higher $\Upsilon$ strengthens gravity, near the stellar center, although it weakens gravity far away from it \cite{SaitoJCAP,SaksteinIJMP}.

This model with all its assumptions is known to be a very good one for a $1.1M_{\odot}$ star at its turnoff point. Now in modified gravity, the radius and luminosity were obtained as functions of the modified gravity parameter $\Upsilon$
\begin{equation}
\log\left(\frac{R}{R_{\odot}}\right) = 0.003 + 0.103\Upsilon - 0.194\Upsilon^{2}~,~~
\log\left(\frac{L}{L_{\odot}}\right)=0.440 - 0.269\Upsilon~.
\label{LRFit}
\end{equation}
Assuming a conservative $\sim 3\%$ error in the measurement of $L$ (for $\Upsilon=0$), the following bound on $\Upsilon$ is obtained
\begin{equation}
-0.05 < \Upsilon < 0.04~.
\label{UpsBoundJCAP}
\end{equation}
It was later shown in \cite{Chowdhury2}, that this stellar model is insensitive to small changes in the metallicity. Therefore, the conclusions above remains unaltered in presence of a small non-zero metallicity, within the ambit of such a model. However, such stellar objects having distinct core-envelope structure are not yet analyzed in the presence of anisotropies, and therefore this topic is left for future studies.

\subsubsection{Intermediate mass stars after core hydrogen-burning ends}
As mentioned in the introduction Sec.\ref{SecStellarPhysics}, a characteristic limit called the SC limit is associated to the post-main sequence evolution of an intermediate mass star. Such a limit describing the upper limit of the helium core mass fraction, is given by the conventional textbook formula $q_{max}\sim 0.37(1/\alpha)^2$ \cite{CarrollOstlie, CoxGiuli, Kippenhahn} in the isotropic Newtonian limit of GR. In this section, we discuss the work of \cite{Chowdhury4}, which analyzes the effect of stellar pressure anisotropy and modified gravity theories on such a limit. 

The stellar model comprises of an isothermal non-degenerate core, surrounded by an envelope in radiative equilibrium. The radiative envelope is modeled by an $n=3$ polytrope, using the Eddington standard model approximation. Using appropriate non-dimensional variables, the associated stellar structure equations are integrated with corresponding boundary conditions and fitting conditions of the homology invariants at the junction, see Eq.\eqref{UVfit}. The numerical technique used here is based largely on \cite{BallTout2012}. More details of the same and associated subtleties, can be found in \cite{Chowdhury4}. The three independent parameters in the model are varied well within their typical admissible ranges; $\alpha$ is varied from $1$ (homogeneous composition) to $2.5$ (completely ionised helium core with completely ionised hydrogen in the envelope), $\Upsilon$ is varied from $-0.5$ to $0.5$, which is the typical range of $\Upsilon$ obtained from the astrophysical probes of modified gravity theories (see \cite{Chowdhury2}), $\tau$ is varied from $-0.01$ to $0.01$ (see \cite{Chowdhury1}).

In the isotropic situation, a quartic fitting formula for the SC limit, $q_{max}=\sum_{x,y}C_{x y}(1/\alpha)^{x}\Upsilon^{y}$, with $x+y\leq4$, is obtained as a function of the modified gravity parameter, from the numerically obtained data points. $C_{xy}$ are numerical coefficients as listed in the following Table \ref{modgravformula}. 
\begin{table}[h]
\centering
\begin{tabular}{|c|c|c|c|c|c|}
	\hline
	$x \backslash y$ & 0 & 1 & 2 & 3 & 4\\
	\hline
	$0$ & $-$ & $-0.018$ & $-0.410$ & $-0.174$ & $0.100$\\
	$1$ & $0.136$ & $0.589$ & $1.50$ & $0.084$ & $-$\\
	$2$ & $-0.544$ & $-2.00$ & $-1.02$ & $-$ & $-$\\
	$3$ & $1.56$ & $1.20$ & $-$ & $-$ & $-$\\
	$4$ & $-0.793$ & $-$ & $-$ & $-$ & $-$\\
	\hline
\end{tabular}
\caption{List of coefficients $C_{xy}$ \protect\cite{Chowdhury4}.}
\label{modgravformula}
\end{table}
Similarly, in the Newtonian limit, a quartic fitting formula for the SC limit, $q_{max}=\sum_{x,y}\bar{C}_{x y}(1/\alpha)^{x}\tau^{y}$, with $x+y \leq 4$, is obtained as a function of the anisotropy parameter, from the numerically obtained data points. $\bar{C}_{xy}$ are numerical coefficients as listed in the following Table \ref{modgravformula4}. 
\begin{table}[h]
	\centering
	\begin{tabular}{|c|c|c|c|c|c|}
		\hline
		$x \backslash y$ & 0 & 1 & 2 & 3 & 4\\
		\hline
		$0$ & $-$ & $-0.061$ & $104$ & $4.77\times10^3$ & $1.57\times10^5$\\
		$1$ & $0.117$ & $-0.662$ & $-442$ & $7.61\times10^3$ & $-$\\
		$2$ & $-0.450$ & $-0.569$ & $400$ & $-$ & $-$\\
		$3$ & $1.42$ & $-1.39$ & $-$ & $-$ & $-$\\
		$4$ & $-0.732$ & $-$ & $-$ & $-$ & $-$\\
		\hline
	\end{tabular}
	\caption{List of coefficients $\bar{C}_{xy}$ \protect\cite{Chowdhury4}.}
	\label{modgravformula4}
\end{table}        

From both these formulas, it is observed that the SC limit decreases for higher values of the modified gravity parameter as well as the anisotropy parameter, which can be physically justified as follows. For higher values of the modified gravity parameter or the anisotropy parameter, the gravitational strength inside the stellar object is lowered, which leads to a decrease in the maximum pressure that can be supported by the isothermal core. Therefore, the limiting core mass fraction is attained at a value lower than the SC limit for the standard case. In \cite{Chowdhury4}, a quartic master formula was obtained as well for the SC limit : $q_{max}=\sum_{x,y,z}C_{x yz}(1/\alpha)^{x}\Upsilon^{y}\tau^{z}$, with $x+y+z \leq 4$, as a function of both $\Upsilon$ and $\tau$. $C_{xyz}$ being numerical coefficients.
\begin{figure*}[h]
	\begin{tabular}{cc}
		\includegraphics[width=58mm]{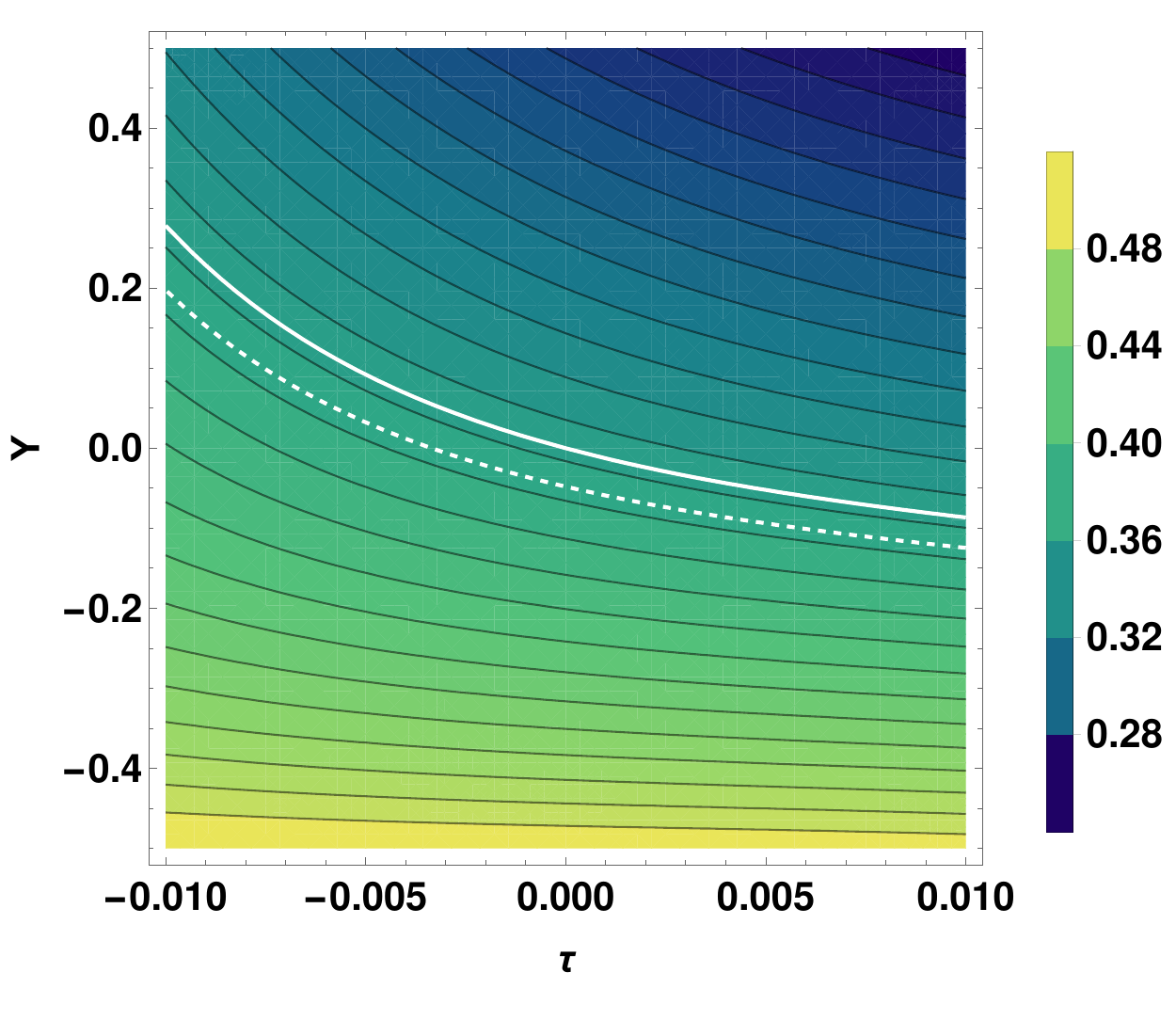} & \includegraphics[width=61mm]{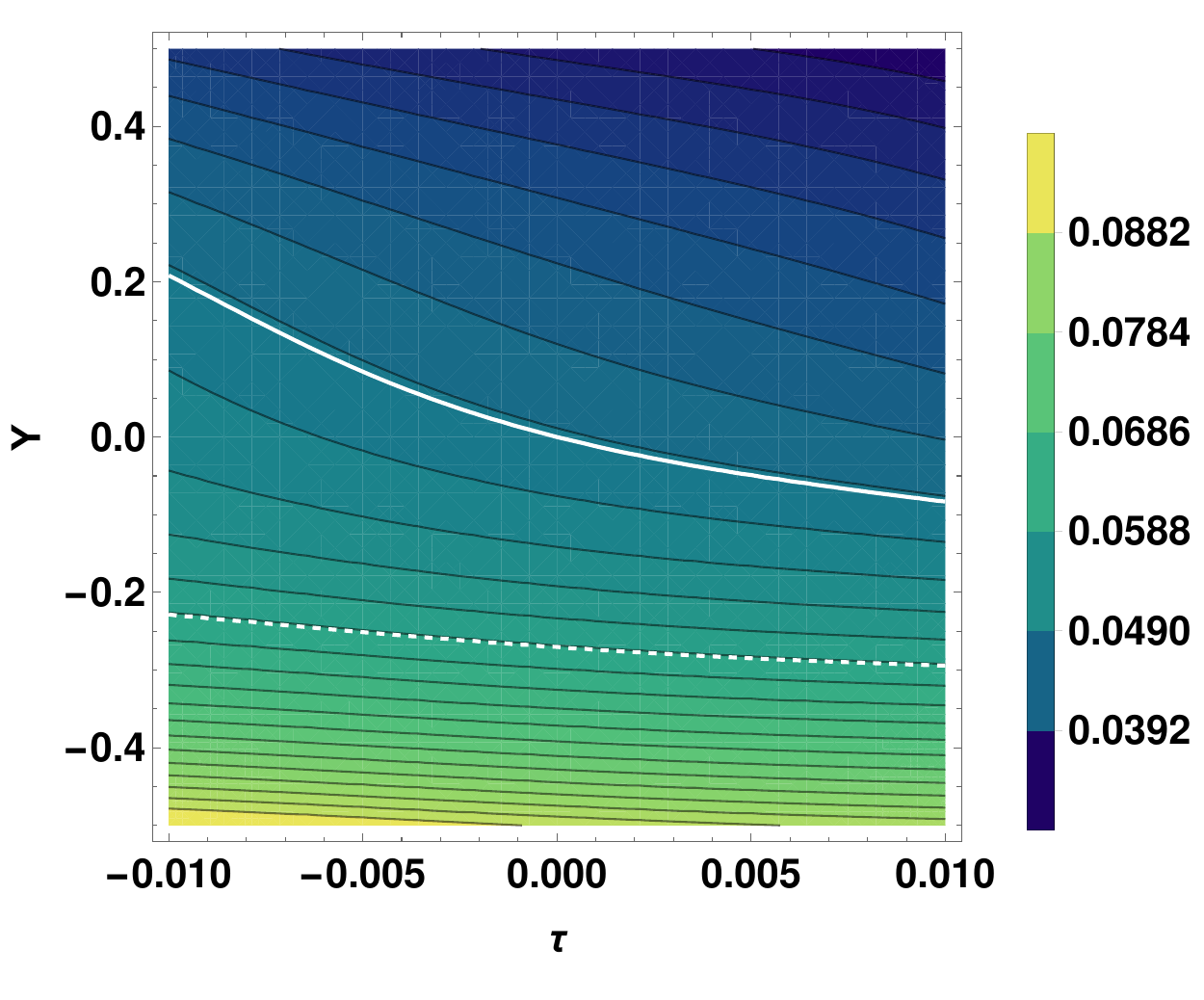} \\
		(a) $\alpha=1$ & (b) $\alpha=2.5$  \\[4pt]
	\end{tabular}
	\caption{Contours of the SC-limit for $\alpha=1$ (a) and $\alpha=2.5$ (b) from 
		the quartic master formula. Any particular 
		contour specifies all admissible tuples of the modified gravity parameter $\Upsilon$ and the non-dimensional 
		anisotropy constant $\tau$, 
		for which one obtains identical SC limit. The white dashed contour corresponds to $q_{max}=0.37(1/\alpha)^2$.
		The white solid contour corresponds to the $\Upsilon=\tau=0$ limit of our master formula \protect\cite{Chowdhury4}.}
	\label{FigSClimitContourPlot}
\end{figure*}

Significant deviations in the value of the SC limit, compared to that in the standard case has been reported in \cite{Chowdhury4}, due to the consideration of stellar pressure anisotropy and modified gravity theories. From Fig.\ref{FigSClimitContourPlot}, it is seen that the white solid contour representing the standard case limit of the quartic master formula differs from the conventional SC limit of $0.37(1/\alpha)^2$ by $\sim 2.8\%$ for $\alpha=1$ and by $\sim 20\%$ for $\alpha=2.5$. One can attribute these differences to the fact that the standard case limit of our quartic master formula includes linear, cubic, and quartic terms in addition to the quadratic term that comprises the conventional formula. For larger values of $\alpha$, these differences are much more pronounced since the contribution from the linear term is enhanced. The most important takeaway message is that, contrary to what is usually done in textbooks, the linear term is just as important as the quadratic term.

An important consequence of such a change in the SC limit was \tred{first} studied in \cite{Maeder}, where stellar rotation was considered. It was shown that an increase in the SC limit results in an observable increase in the number of stars in their shell hydrogen-burning phase, near the main-sequence. \tred{Similar physical consequence has also been established and quantified in \cite{Chowdhury4}. In the isotropic case, for negative values of $\Upsilon$, the authors estimated an increase in the shell hydrogen-burning lifetime compared to its value in the standard case. For positive values of $\Upsilon$, however, the authors could arrive at upper bounds on the modified gravity parameter for different values of $\alpha$. Uniform stellar rotation, however, leads to a further small decrease in the SC limit \cite{Maeder} and, therefore, should make the bounds sharper.}

\tred{The significance of the modified gravity and anisotropy lies in the fact that they allow for modifications in the SC limit while remaining within the framework of classical dynamics. For example, inside the partially degenerate core of a star with mass $M \sim 1.4M_{\odot}$, degeneracy pressure, which results from quantum effects, can play a crucial role in supporting a higher value of the envelope pressure. As we have seen, the modified gravity and anisotropy can mimic this effect to a small degree at the classical level, i.e., with an isothermal non-degenerate core. Similarly for stars with $M \sim 8M_{\odot}$, such modifications imply that a star that would have undergone contraction according to the standard framework may not do so in the modified scenario and vice versa.}

\subsubsection{SGB to AGB}
The post-main sequence phase discussed above is followed by the SGB, RGB, and going up along the AGB, as the star evolves with time. However, these phases, specially the SGB and the AGB, have not been studied widely in the framework of surviving-DHOST theories. Although some works on modified gravity have appeared for the RGB phase \cite{PhilipChangRGBChameleon, NajafiRGBChameleon}, such works are within the framework of chameleon theories \cite{KhouryWeltmanChameleon,MotaBaroChameleon}, which have been reported to be heavily constrained through main-sequence modeling \cite{SaksteinConstrainChameleon1} using Modules for Experiments in Stellar Astrophysics (MESA) \cite{MESA1,MESA2}, and different distance indicators \cite{SaksteinConstrainChameleon2,SaksteinConstrainChameleon3}. 

Further, according to the suggestion of \cite{PhilipChangRGBChameleon}, the surviving-DHOST theories might not alter the structure of red-giant stars, but the RGB has not yet been studied in the context of surviving-DHOST theories, and is therefore a potential open area of research. Another possible avenue for future research is AGB mass loss due to a companion star \cite{Malfait1,Malfait2} within the framework of modified gravity theories. Such mass loss during the AGB phase, has important implications on the final fate of the star having ZAMS mass between $4M_{\odot}$ and $8M_{\odot}$. The AGB mass loss prevents the CO core of such stars from undergoing catastrophic core collapse eventually \cite{CarrollOstlie}. The CO core of stars having ZAMS mass below $4M_{\odot}$ however, develop degeneracy, which prevents the central temperature from attaining appreciable values to ignite subsequent nuclear burning reactions irrespective of whether it undergoes mass loss during AGB.

\subsubsection{White dwarfs}
The low and intermediate mass stars, i.e., stars having masses below $8M_{\odot}$, eventually becomes white dwarfs after the AGB phase, as mentioned in the introduction, Sec.\ref{SecStellarPhysics}. The maximum mass of such a white dwarf in the standard situation has been shown to be $\sim 1.4M_{\odot}$ \cite{ChandraWD1,ChandraWD2,ChandraWD3}, by considering a $n=3$ polytrope, which models the relativistic degenerate electron gas inside such highly dense WDs. This is the famous Chandrasekhar limit. 

\begin{figure}[h]
	\centering
	\includegraphics[width=0.9\linewidth]{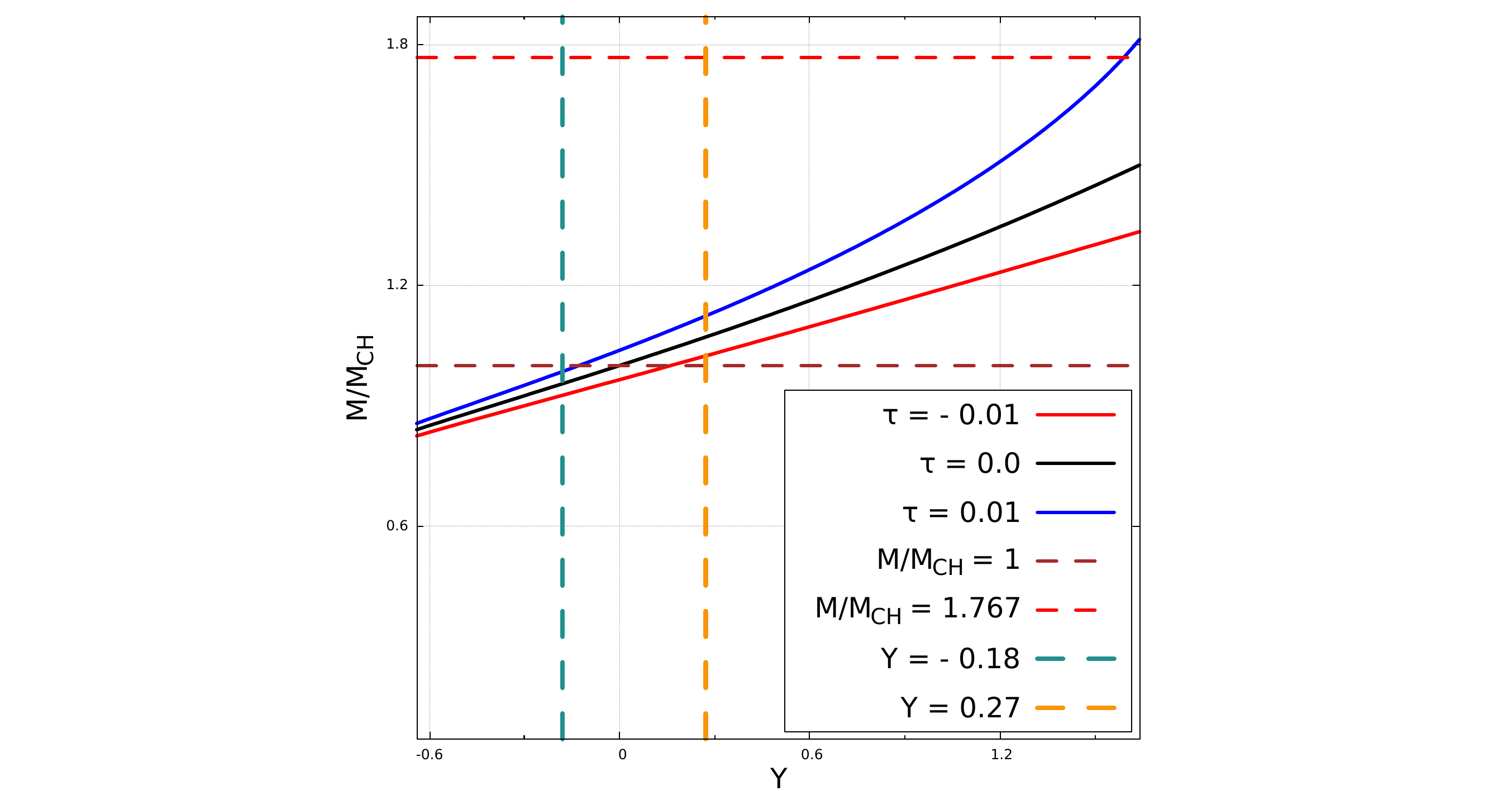}
	\caption{Mass of WD (in units of Chandrasekhar limit $M_{\rm CH}$) as a function of $\Upsilon$ for different $\tau$ values \protect\cite{Chowdhury1}.}
	\label{FigWDmass}
\end{figure}

However, it was shown in \cite{Chowdhury1} that if stellar pressure anisotropy is considered within the framework of modified gravity theories, then WDs can have masses beyond the Chandrasekhar limit. In \cite{Chowdhury1}, a polytropic EOS with $n=3$, was considered. For a polytropic model having stellar pressure anisotropy in modified gravity theories, the stellar density profile is obtained from the solution of the MLEE Eq.\eqref{MLEE}. From the solution one then obtains the mass of the polytrope as 
\begin{equation}
M = 4\pi\int_0^R \rho(r) r^2 dr = -4\pi r_c^3 \rho_c \xi_R^2\theta'(\xi_R)~,
\label{MCh}
\end{equation}
where $\xi_R$ is the non-dimensional stellar radius. Note that although the expression for mass does not contain the anisotropy parameter $\tau$ or the modified gravity parameter $\Upsilon$, their effect upon the mass is implicit through the solution $\theta$ of the MLEE Eq.\eqref{MLEE}, which explicitly contains these parameters. The value of $M$ for $\Upsilon=0=\tau$ corresponds to the Chandrasekhar mass limit; we will denote it by $M_{\rm CH}$. The effect of anisotropy and modified gravity parameters on the WD mass can be see from Fig.\ref{FigWDmass}, where the ratio of the WD mass $M$ to $M_{\rm CH}$ is plotted.

It is observed that both higher values of the anisotropy parameter as well as the modified gravity parameter results in the increase of WD mass. The reason for such behavior can be justified as follows. The maximum mass of a WD corresponds to that point at which gravity overcomes
electron degeneracy pressure. We know that higher values of the modified gravity and anisotropy parameters reduce the effective gravitational strength. Therefore, the object needs more mass in order to increase the effective gravitational strength and overcome the electron degeneracy pressure. This explains the observed super-Chandrasekhar white dwarfs, even without the assumption of high magnetic fields in the core.

There are several works along this line, where white dwarfs are studied in the context of modified gravity theories, although mostly in isotropic situations. In \cite{BaniDebFred}, the authors studied local anisotropies in highly magnetized WDs. One can possibly extend this work to include modified gravity theories as well. Several modified gravity theories have been constrained through studies on WDs. For example, while \cite{JainWD} obtained strong constraints on the beyond Horndeski parameter from the mass-radius relation of WDs, \cite{SrimantaWD} constrained the parameters corresponding to other modified gravity theories from such mass-radius relations. \tred{In \cite{KalitaBaniGWfR}, the authors estimated the difference in the amplitudes of the relevant polarization modes of the gravitational waves sourced by regular white dwarfs ($M \lesssim 1.4 M_{\odot}$) and the peculiar ones ($M \gtrsim 2.8 M_{\odot}$) and thus proposed direct evidence of detecting the peculiar white dwarfs, which can thus constrain the underlying modified gravity theory.} Other interesting features like the cooling process in WDs \cite{KalitaWojnarCoolingWD}, crystallization inside WDs \cite{KalitaWojnarCrystalWD}, and the stability of WDs \cite{LupamudraWojnarStabilityWD} in the context of Palatini $f(R)$ gravity have been extensively studied. Such studies can be carried out in the framework of surviving-DHOST theories as well. Several important questions pertaining to WDs in modified gravity theories were posed in \cite{WojnarWD}. \tred{Other recent works on the properties and stability of WDs in modified gravity include \cite{UtamiSulakso,Carvalho}, while the authors of \cite{Chowdhury6} have very recently studied the tidal disruption of WDs in modified gravity theory using smoothed particle hydrodynamics (SPH)}. Interestingly in most of these works, the EOS being used was not considered to be modified due to modified gravity theories. However, it was noted very recently in \cite{WojnarEOSWD} that the EOS describing a Fermi gas indeed gets modified depending upon the modifications in the gravitational interactions.

\section{Stellar rotation}
\label{SecRotation}
In the previous Sec.\ref{SecModAnisotropy}, we discussed stellar pressure anisotropy in situations where spherical symmetry is retained. However, if the anisotropy originates due to stellar rotation, then this symmetry is broken in reality, and the assumption of spherical symmetry needs to be relaxed. 

In this section we will first discuss the interesting effects of rapid uniform stellar rotation on VLM objects in the Newtonian framework \cite{Chowdhury3}. We will then outline a recent theorem which provides the analytic formulation of slowly rotating stellar objects or substellar objects within the framework of modified gravity theories \cite{Chowdhury5}.

\subsection{Rapidly rotating VLM objects}

In the recent past, astrophysical observations have revealed some BDs to rotate rapidly \cite{Route,Clarke,Metchev,Williams,Tannock}, having a considerably shorter rotational period than Jupiter, which has a rotation period of $\sim 10$ hours. The smallest rotation period of $1.08$ hours of BDs is indicated as a lower limit on such rotation periods, 
based on the observed clustering of such objects having similar rotation periods \cite{Tannock}.

Interestingly such constraints on rotations of VLM objects have been deduced from theoretical modeling in \cite{Chowdhury3}. The authors considered the polytropic EOS for VLM objects, as defined in Sec.\ref{SecVLM}. Along with it the Poisson equation
\begin{equation}
\nabla^2 \Phi = 4\pi G_N \rho~,
\end{equation}
and the Euler equation corresponding to momentum conservation
\begin{equation}
\label{Euler}
\rho \frac{\partial v^i}{\partial t} + \rho v^j \frac{\partial v^i}{\partial x^j} = 
-\frac{\partial P}{\partial x^i} - \rho \frac{\partial \Phi}{\partial x^i}~,
\end{equation}
are used. In the above Eq.\eqref{Euler}, $v^i=\Omega\{x^3,0,-x^1\}$ corresponds to the velocity field of the rotating object, in a Cartesian coordinate system. For a given mass $M$, rotating with a given angular speed 
$\Omega$, the equilibrium configuration of the VLM object at a particular time of its evolution, parameterized by the degeneracy parameter $\eta$, is obtained from a self iterating scheme involving the Poisson equation and the Euler Equation \cite{Ishii}. Corresponding to the obtained equilibrium configuration, the $L_{\rm HB}$ and $L_{\rm s}$ are calculated, by incorporating the thermodynamics appropriate for a VLM object, see Sec.\ref{SecVLM}. The condition for stable hydrogen-burning is therefore checked by evolving VLM objects of differing masses for every given $\Omega$. 

Performing the above numerical analysis, the authors of \cite{Chowdhury3} found that while $M_{\rm mmhb}$ increases with increase in $\Omega$, there exists a maximum mass of stable hydrogen-burning, called $M_{\rm max}$, which decreases with increase in $\Omega$. Importantly, this shows the existence of overmassive BDs, i.e., BDs having masses higher than 
$\sim 0.08M_\odot$ corresponding to the non-rotating case, without the consideration of accretion effects, which was studied in \cite{ForbesLoeb}.
\begin{figure}[h]
	\centering
	\includegraphics[width=0.65\linewidth]{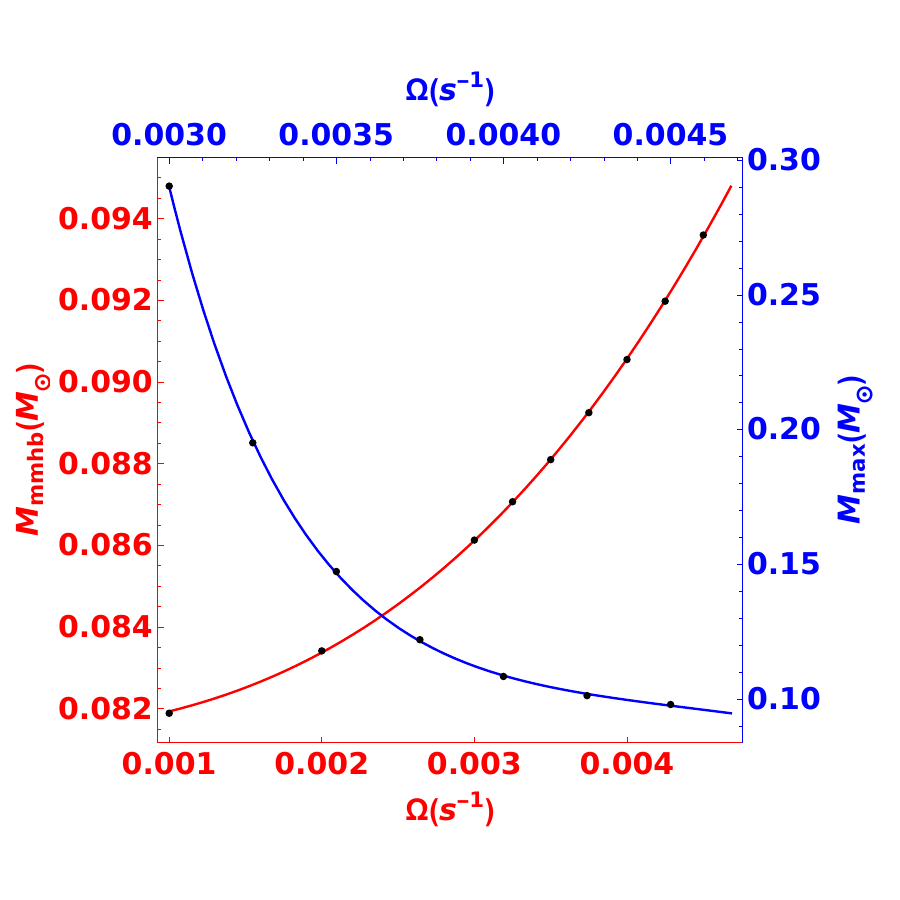}
	\caption{The red curve corresponds to $M_{\rm mmhb}(\Omega)$. The blue curve corresponds to $M_{\rm max}(\Omega)$. Each y-axis is color coded to the data \protect\cite{Chowdhury3}.}
	\label{MMHB_MMAX}
\end{figure}

While the increase in $M_{\rm mmhb}$ due to increase in $\Omega$ can be attributed to the reduction in effective gravitational strength inside the stellar material due to the centrifugal forces, the existence of $M_{\rm max}$ needs careful explanation. A given stellar rotation $\Omega$ can be sustained by a VLM object only if it possesses a central density higher than a certain critical value, called the critical density $\rho_{\rm crit}$. Moreover, in non-rotating cases, if a given VLM object at some point of its evolution attains higher hydrogen-burning luminosity compared to its surface luminosity, then it can undergo internal readjustments by expanding and lowering its central density and temperature and thus the hydrogen-burning luminosity until the stable hydrogen-burning condition is retained. However, in presence of rotation $\Omega$, it is found that  VLM objects having masses higher than $M_{\rm max}$, cannot undergo such internal readjustments, since upon doing so it will eventually reduce its central density below the critical value $\rho_{\rm crit}$, required to sustain the rotation. Hence, such VLM objects cannot attain stable hydrogen-burning conditions while maintaining the given rotation. For a given $\Omega$, the $M_{\rm max}$ corresponds to the equilibrium configuration with critical central density, for which the stable hydrogen-burning condition is attained; it is called the {\it critical} configuration. Now, with increase in $\Omega$, the 
$\rho_{\rm crit}$ increases, thereby increasing the hydrogen-burning luminosity of the corresponding critical configuration. Thus a lower mass is required for the critical configuration to attain stable hydrogen-burning condition. 

The corresponding formulas for the $M_{\rm mmhb}(\Omega)$ and $M_{\rm max}(\Omega)$ obtained in \cite{Chowdhury3}, are as follows
\begin{align}
& M_{\rm mmhb}(\Omega)= 0.0814 + 0.2302 \Omega + 245.58 \Omega^2 + 67646 \Omega^3~, \label{MMHBformula} \\
& M_{\rm max}(\Omega) =  -1.2621 + 0.0176\Omega^{-1} - 7.7873 \times 10^{-5}\Omega^{-2} + 1.1648 \times 10^{-7}\Omega^{-3}~,
\label{MMAXformula}
\end{align}
where $\Omega$ is in $s^{-1}$ and the formula gives $M_{\rm mmhb}(\Omega)$ as well as $M_{\rm max}(\Omega)$ in units of the solar mass $M_{\odot}$. For a given $\Omega$, the VLM objects having masses within the transition mass range ($M_{\rm mmhb}(\Omega) \leq M \leq M_{\rm max}(\Omega)$) attains stable hydrogen-burning condition and reaches the main-sequence. This transition mass range decreases with increase in $\Omega$, and eventually tapers to zero at a particular value $\Omega_{\rm max}\sim 0.0047~s^{-1}$ (rotation period of $\sim 22$ minutes). Beyond this maximum value of $\Omega=\Omega_{\rm max}$, no model solutions exist, and thus a VLM object does not reach the main sequence, within the ambit of the theoretical model considered here. In \cite{Chowdhury3}, a master formula for the stellar luminosity of the VLM object when it reaches the main-sequence phase was obtained as a function of both the stellar mass and the uniform angular rotation speed,
\begin{equation}
\tilde{L}_{HB}(M,\Omega)/L_{\odot} = \sum_{\alpha,\beta}\tilde{C}_{\alpha\beta}(M/M_{\odot})^{\alpha}(\Omega/s^{-1})^{\beta}~,
\end{equation}
with the coefficients $\tilde{C}_{\alpha\beta}$ being listed in Table \ref{TableL}. 
\begin{table}[h!]
	\centering
	\renewcommand{\arraystretch}{1.3}
	\label{TableL}
	\resizebox{\columnwidth}{!}{\begin{tabular}{|c| c| c| c| c| c| c| c| }\hline
			$\alpha \backslash \beta$ &~0 &1 &2 &3 &4 &5 &6 \\ \hline
			$0$& $24.203$& $1265.4$& $95238$& $1.2927\times 10^{6}$& $6.7610\times10^6$& $7.3972\times10^9$& $1.114\times10^{11}$ \\
			$1$& $-1577.2$& $-70298$& $-4.1388\times10^6$& $-3.3679\times10^7$& $-8.8127\times10^8$& $-1.0444\times10^{11}$& $-$\\
			$2$& $42761$& $1.5598\times10^6$& $6.6859\times10^7$& $2.8294\times10^8$& $1.0488\times10^{10}$& $-$& $-$ \\
			$3$& $-6.1736\times10^5$& $-1.7279\times10^7$& $-4.7498\times10^8$& $-7.9453\times10^8$& $-$& $-$& $-$ \\
			$4$& $5.0065\times10^6$& $9.5552\times10^7$& $1.2490\times10^9$& $-$& $-$& $-$& $-$ \\
			$5$& $-2.1626\times10^7$& $-2.1102\times10^8$& $-$& $-$& $-$& $-$& $-$ \\
			$6$& $3.8887\times10^7$& $-$& $-$& $-$& $-$& $-$& $-$ \\
			\hline
	\end{tabular}}
	\caption{List of coefficients $\tilde{C}_{\alpha\beta}$ \protect\cite{Chowdhury3}.}
\end{table}

Although rapidly rotating neutron stars have been investigated in the context of certain modified gravity theories \cite{DonevaNSRot,YazadjievNSRot}, such studies have not yet been performed in the context of VLM objects. Neutron stars being highly compact, one needs a metric formalism, while VLM objects are studied in the Newtonian limit, using the Poisson equation. Therefore, one can in principle extend the results of \cite{Chowdhury3} by including the appropriate correction term in the Poisson equation due to the modified gravity theories in the Newtonian limit.

\subsection{Slow rotation within modified gravity theories}

Slowly rotating compact objects have been studied in modified gravity theories in the relativistic framework \cite{Kleihaus,Silva,Komatsu,Paschalidis}. In such scenarios, a fully numerical approach is implemented by considering a metric corresponding to the star. However, less compact stars in the Newtonian limit of modified gravity theories have not been studied widely. In this section we discuss the work of \cite{Chowdhury5}, where the analytical formalism to deal with slow stellar rotation in the Newtonian limit of modified gravity theories has been provided.

In the presence of rotation, the pressure balance equations
\begin{align}\label{HECrot1}
\frac{\partial P}{\partial r} &= -\rho\frac{\partial \Phi}{\partial r} + \rho\Omega^2r(1-\mu^2)~,\\
\frac{\partial P}{\partial \mu} &= -\rho\frac{\partial \Phi}{\partial \mu} - \rho\Omega^2r^2\mu~,
\label{HECrot2}
\end{align}
with the rotation induced centrifugal terms need to be considered \cite{Chowdhury5}. In the above equations, $P$ corresponds to the net fluid pressure with $\mu(=\cos\vartheta)$ being the angular coordinate. The stellar rotation $\Omega$ can be uniform as well as non-uniform. While Eqs.\eqref{HECrot1} and \eqref{HECrot2} do not include any explicit terms for modified gravity, such effects are captured by the potential $\Phi$, through the modified Poisson equation
\cite{EiBI},\cite{WojnarNAV},\cite{Pritam1}
\begin{equation}
\nabla^2\Phi = 4\pi G_N \rho + L\Phi_{mod}(r,\mu)~.
\label{ModPoisson}
\end{equation}
In Eq.\eqref{ModPoisson}, $L\Phi_{mod}(r,\mu)$ is a generic correction term due to the modified gravity theories being considered in the presence of rotation. At this point it needs to be emphasized that the pressure, density, correction term due to modified gravity theories, as well as the potential considered in the above equations are functions of both the radial and the angular coordinates due to the rotation-induced asymmetry.

Now, according to the proposition put forward in \cite{Chowdhury5}, if $\Theta(\xi,\mu)$ is the solution to the rotating polytrope of index $n$ with uniform angular speed $\Omega$, then the corresponding Lane-Emden equation in modified gravity is given by
\begin{equation}
\frac{1}{\xi^2}\frac{\partial}{\partial \xi}\Big(\xi^2 \frac{\partial \Theta}{\partial \xi}\Big) + \frac{1}{\xi^2}\frac{\partial}{\partial \mu}\Big((1-\mu^2) \frac{\partial \Theta}{\partial \mu}\Big) \nonumber
= v + g_{mod}(\xi,\mu)-\Theta^n~,
\label{modLEE}
\end{equation}
where $v=\Omega^2/2\pi G_N \rho_c$ is a dimensionless parameter which is a measure of the outward centrifugal force compared to the self-gravity of the rotating polytrope, while $g_{mod}(\xi,\mu)=-L\Phi_{mod}/4\pi G_N \rho_c$ is the dimensionless modification term depending on a given theory of gravity in general. With $\theta(\xi)$ denoting the solution to the non-rotating polytrope in modified gravity, such a proposition is then supplemented by the following theorem,

\textbf{Theorem:} ``If $g_{mod}(\xi,\mu)$ can be expanded in terms of the Legendre functions $P_l(\mu)$'s as
\begin{equation}\label{TheoremEq1}
g_{mod}(\xi,\mu) = g_{mod0}(\xi) +v\Bigg\{\bar{\tilde{g}}_{mod}(\xi)P_0(\mu) + \sum_{j=1}^{\infty} \bar{\bar{\tilde{g}}}_{modj}(\xi)P_j(\mu)\Bigg\}~,
\end{equation}
where $g_{mod0}(\xi)$ is the non-rotating part, with $\bar{\tilde{g}}_{mod}(\xi)$, and $\bar{\bar{\tilde{g}}}_{modj}(\xi)$ being the rotation induced ones, then the solution $\Theta$ of the modified Lane-Emden equation in presence of slow rotation is
\begin{equation}
\Theta(\xi,\mu) =\theta(\xi) + v\Big[\psi_0(\xi) + \tred{\bar{A}}_2\psi_2(\xi)P_2(\mu)\Big]\nonumber~,
\end{equation}
where $\psi_0$ and $\psi_2$ satisfy the following equations:
\begin{equation}
\frac{1}{\xi^2}\frac{d}{d \xi}\Big(\xi^2 \frac{d \psi_0}{d \xi}\Big) = - n\theta^{n-1}\psi_0 + 1 + \bar{\tilde{g}}_{mod}(\xi)~,
\end{equation}
\begin{equation}
\frac{1}{\xi^2}\frac{d}{d \xi}\Big(\xi^2 \frac{d \psi_2}{d \xi}\Big) = \Big(\frac{6}{\xi^2}-n\theta^{n-1}\Big)\psi_2 + \frac{\bar{\bar{\tilde{g}}}_{mod2}(\xi)}{\tred{\bar{A}}_2}~,
\end{equation}
where
\begin{equation}
\tred{\bar{A}}_2= - \frac{5}{6}\frac{\xi_1^2}{[3\psi_2(\xi_1)+\xi_1 \psi_2^{'}(\xi_1)]}~,
\label{A2}
\end{equation}
with $\xi_1$ being the first zero of $\theta(\xi)$ while a $'$ denotes a derivative with respect to $\xi$."

The above theorem is in fact very general and includes several important classes of modified gravity theories \cite{Chowdhury5}. For example it was shown that the above analytical formalism can be carried out in beyond Horndeski theories, metric-affine Palatini $f(R)$ and EiBI gravity models.

\section{\tred{Strong field effects and Neutron Stars}}
\label{SecStrongField}

\tred{At this point, it is crucial to note that we have confined our interest in the Newtonian limit, i.e., weak field approximation, of the gravity theories. Such an approximation is valid and widely used in studying the classes of stars -- very low mass objects, low mass stars during the shell hydrogen-burning phase, and intermediate mass stars, that we have considered in the previous sections. However, the corrections due to strong gravitational field need to be incorporated for compact objects like neutron stars and higher mass white dwarfs.} 
	
\tred{The requirement for the strong gravity corrections can be estimated from the so-called compactness parameter $\sim 2G_{N}M/c^2R$, whose value is $0.001$ for the WDs, $0.3$ for the neutron stars, and $1$ for black holes \cite{Shapiro1983}. This parameter actually comes from the Tolman-Oppenheimer-Volkov (TOV) equation, i.e., the pressure balance equation in GR, where $(1-2G_{N}M(r)/c^2r)^{-1}$ appears as a multiplicative factor which stands for the strong gravity corrections \cite{SilbarReddy}. A larger value of the compactness parameter, thus, enhances the strength of gravity. Therefore, we see from the above illustrative values that strong gravity corrections are indeed important for neutron stars. For other kinds of stars discussed in this paper, such corrections should be negligible. However, it can be an interesting exercise to quantify the deviations in stellar observables due to modified gravity and anisotropy compared to those derived from the relativistic and strong field corrections in such low dense stellar objects.}

\subsection{\tred{Strong field effects}}

\tred{Although in GR, it is well known that strong field effects can be ignored for white dwarf stars, it need not be the case in beyond-Horndeski theories where the strength of gravity gets modified (see \cite{BabichevKoyama}). The authors of this paper showed that strong field corrections indeed result in a different behavior of the maximal mass of white dwarfs compared to the non-relativistic case, although the mass-radius relation remains invariant for low mass white dwarfs. Therefore, the bound obtained in \cite{JainWD} from the mass-radius relationship of low mass WDs, in the Newtonian limit, should, in principle, remain unchanged. However, the maximal mass of WDs obtained in \cite{Chowdhury1} and the bounds obtained from them \cite{JainWD} should, in principle, alter when the strong field effects are considered.}

\subsection{\tred{Neutron stars}}

\tred{For completeness, we finally discuss how the modified gravity theories affect the neutron stars and draw a comparison between the bounds obtained from them and the ones we discussed in this article from other classes of stars. Neutron stars in beyond-Horndeski theories have been studied in \cite{SaksteinStrongFieldTest,BoumazaLanglois}. The authors have found the mass-radius relationship of neutron stars to deviate from those derived in Einstein's theory of general relativity -- larger values of the modified gravity parameter $\Upsilon$ correspond to lower maximum masses and smaller radii for fixed mass (see \cite{BabichevKoyama}). From such a relationship, they could constrain the modified gravity parameter to $\Upsilon \gtrsim -0.03$, below which the maximal mass of neutron stars exceeds the observed maximal value of $\sim 2M_{\odot}$. However, the obtained constraint is heavily dependent upon the EOSs being considered because the mass-radius relationship of neutron stars is found to be degenerate with the equation of state (EOS), which is highly uncertain. The aforementioned bound is comparable to the most stringent bound $\Upsilon \gtrsim -0.05$ obtained from low mass population II star in post-main-sequence evolution (see Sec.\ref{SectionpopII}). The lower bounds discussed in this review article from the other types of stars are less constrained than this. Neutron stars have also been studied in other modified gravity theories like $f(R)$ gravity \cite{CapozzielloNSfR,KaseTsujikawaNSfRSTT} \tteal{and teleparallel gravity \cite{NashedCompactTele}}. Anisotropy in highly magnetized neutron stars in GR has been studied in \cite{BaniDebFred2}, which one can, in principle, extend into modified gravity theories. \tteal{Anisotropic neutron stars in Krori-Barua spacetime have been studied in \cite{NashedNSKB}.}}

\section{Discussion and Conclusion}

\tred{The modified gravity theories are successful alternatives to dark energy in explaining the late-time universe, as mentioned earlier. While these modified theories have been extensively investigated in the cosmological aspects, their implications on stellar structure and evolution have only gained interest in the last decade. The motivation behind such studies is simple: besides the cosmological tests, these modified gravity theories have to pass the various astrophysical tests as well. That is why studying stellar objects in modified gravity theories and confronting them against the observations is an integral part of the attempts to find a generalized theory of gravity. Most of these studies were restricted to compact objects like neutron stars, with the less compact objects remaining a poorly studied branch. However, only recently, there has been a significant amount of studies on the less compact objects like very low mass objects, main sequence stars, post main-sequence stars in the Newtonian limit of modified gravity theories.}
 
In this review, we have presented a broad overview and the current status of the implications of DHOST theories \tred{on stellar physics, in the Newtonian limit,} after GW170817. This was done in the presence of local pressure anisotropies, retaining the spherical symmetry as well as in presence of stellar rotations, which breaks the spherical symmetry. We have systematically introduced the class of DHOST theories, and 
the partial screening of the scalar degree of freedom inside the anisotropic stellar and substellar objects. Through different stellar modeling appropriate for such objects at different points of their evolution, we showed how one obtains the constraints on theory parameters, by comparing the modified predictions with observational data of stellar parameters like mass, radius and luminosity. We have also shown how a limiting rotation speed of VLM objects can be obtained from theoretical modeling, in the Newtonian framework, and how slow rotation can be analytically formulated in the context of any modified gravity theory in general. Although planets have been extensively studied by \cite{WojnarGiantPlanet,WojnarTerrestrial1,WojnarExo,WojnarJupiter,WojnarTerrestrial2} in the context of metric-affine theories, they have not been widely studied in the framework of DHOST theories. \tred{The main results derived from the study of modified gravity and anisotropy in different stellar and substellar objects, discussed in this review, are highlighted below:}

\begin{enumerate}
	\item \tred{In polytropic models, a generic theoretical upper bound $\Upsilon_{\rm max}$ on the modified gravity parameter is obtained, which depends upon the anisotropy parameter $\tau$ and the class of star considered, i.e., polytropic index $n$ \cite{Chowdhury1}. Such an upper bound has not been reported prior to this particular study.}
	\item \tred{In \cite{Chowdhury1}, it is shown that the effects of anisotropy is degenerate with those due to modified gravity in terms of strengthening or weakening gravity inside VLM objects, depending upon the sign of the corresponding parameters $\tau$ and $\Upsilon$. This indicates that the bounds on $\Upsilon$ obtained earlier in the isotropic context, for example, in \cite{SaksteinPRL}, will change in presence of anisotropy.}
	\item \tred{In \cite{Chowdhury2}, the authors have taken the first step towards analytically studying the effects of modified gravity theories inside a low mass, composite stellar object that has a distinct core-envelope structure with an intermediate hydrogen-burning shell. While their study indicates an overall weakening of gravity with an increase in $\Upsilon$, which is consistent with previous studies in polytropic models, it also indicates a non-trivial strengthening of gravity close to the center. The authors also obtained a stringent bound on $\Upsilon$ from the luminosity formula derived as a function of $\Upsilon$.}
	\item \tred{The authors of \cite{Chowdhury4}, for the first time, studied the implications of both anisotropy and modified gravity on the SC limit, which corresponds to the maximum core mass fraction that can support the overlying pressure of the envelope in an intermediate mass star. They derived a three parameter $\{\alpha,\Upsilon,\tau\}$ master formula for the same, with $\alpha$ being the ratio of the mean molecular weight of the core to that of the envelope. A change in the SC limit leads to a change in the time spent by a star in the shell hydrogen-burning phase, from which the authors derived upper bounds on both the modified gravity and anisotropy parameters. It was also shown that in the standard case, the linear term in $1/\alpha$ is as significant as the quadratic term and should not be left out of the conventional formula for the SC limit, which is taken to be a homogeneous quadratic function in $1/\alpha$.}
	\item \tred{It is shown in \cite{Chowdhury1}, that the WD mass exceeds the Chandrasekhar mass limit of $\sim 1.4 M_{\odot}$, when modified gravity and anisotropy are considered. Such a result is extremely important in the light of recent discoveries of super-luminous type Ia supernovae, which are indicative of super-Chandrasekhar white dwarfs having mass exceeding $ 1.4 M_{\odot}$.}
	\item \tred{In \cite{Chowdhury5}, an analytic formalism to derive the equilibrium density profile of a slowly rotating polytrope in any modified gravity theories, in general, is presented. Such a study is a very general one and has been shown to be applicable in various modified gravity theories. In \cite{Chowdhury3}, rapid rotation in VLM objects is studied in the Newtonian limit of GR. The authors obtained a novel theoretical upper bound in the uniform angular rotation speed, beyond which a VLM object cannot reach the main-sequence phase within the ambit of the model considered. The effect of rapid rotation in such objects within the framework of modified gravity theories is left for future study.}
	\item \tred{Finally, the strong field effects on the stellar observables of compact objects in modified gravity theories are discussed, and the results are compared with the ones derived from the less compact objects, which are studied within the Newtonian limit of modified gravity theories. It is found that in the case of high mass WDs, the strong field effects in modified gravity become important. The bounds obtained from the neutron stars are shown to be comparable with some of those derived from the less compact objects in the weak field limit of modified gravity theories.}
\end{enumerate}

\tred{The various constraints on the modified gravity theories derived from stellar astrophysical context naturally lead to the question regarding the fate of these theories, i.e., whether the heavily constrained modified gravity theories are ruled out from being possible alternatives to GR. It is commonly believed that if the parameters characterizing a particular class of modified gravity theories are constrained to take up values very close to zero, then it indicates that such theories are as good as GR and, therefore, deemed redundant. However, in the case of STTs, at least, there are a few subtleties involved with such a simple-minded deduction, which needs to be addressed. Firstly, it has to be noted that the parameters of the class of theories being constrained from the stellar observables appear at smaller scales, i.e., astrophysical scales. For example, in quadratic DHOST theories discussed here, the parameters $\Upsilon_1, \Upsilon_2,$ and $\Upsilon_3$ appear due to the partial screening inside an astrophysical object within the Vainshtein radius $r_{V}$. Therefore, the constraints on such parameters, naturally, puts restriction on the form of only those free functions (for example, $f$ and $A_3$ in quadratic DHOST) which appear in the definition of those parameters at smaller scales, and thus on the possible number of modified gravity theories constituting that particular class. On the other hand, those free functions, if any, which do not appear in the definition of these parameters of that particular class of theories at smaller scales, however, will constitute the degrees of freedom at large scales, i.e., cosmological scales. Secondly, it is possible that even if all the parameters, appearing at smaller scales, are tightly constrained from astrophysical observations, all the free functions of the class of theories need not be constrained. For example, $f$ can still be non-zero in the quadratic DHOST theories discussed here, even in the case $\Upsilon_1, \Upsilon_2$ and $\Upsilon_3$ are all zero (see Eq. \eqref{UpsilonExpression}), i.e., one will still have the freedom over the choice of the non-minimal coupling between scalar field and gravity, which can, however, be tuned from studies at larger scales. In the case of quadratic DHOST theories, therefore, we see that tight constraints on the theory parameters do indicate that the entire class reduces to the class of traditional STTs in the absence of quadratic terms. However, in general, the bounds on the modified gravity parameters, obtained theoretically as well as from stellar observables, need not necessarily rule out a particular class of modified gravity theories completely; they indicate the tuning of the free functions and, thus, a reduction in the number of possible theories within that class. Another interesting aspect to note from the discussion in Sec.\ref{Secpartialbreaking} is that the parameters of DHOST theories appearing at small scales are time-dependent. Their evolution with time will be governed by the cosmological evolution of the background $\phi_0$ \cite{HorndeskiBeyondReview}. Therefore, the variations in these parameters within the timescales involved with the astrophysical studies in modified gravity, like the evolution of stellar/substellar objects or the tidal disruption events, need to be small enough to approximate the theory parameters to be spatially and temporally constant.}

It was very recently shown in \cite{CreminelliLewandoeski}, that the gravitational waves decay into scalar field fluctuations in DHOST theories, barring a specific subclass of such theories. The screening mechanism in this particular subclass is found by \cite{CrisostomiVernizzi, HiranoKobayashi} to be distinct from that in the surviving-DHOST theories discussed here. Therefore, constraining such a subclass of DHOST theories from observational data should be an interesting exercise.

Studying stellar and substellar objects in modified gravity theories are critical for understanding gravitational interactions, even more so considering the large amount of observational data that will be available in the near future. Missions like GAIA \cite{GAIA} and the James Webb Space Telescope \cite{JamesWebb} are expected to provide precise observational data soon. Therefore, all predictions and constraints on modified gravity theories should be possible to verify up to high accuracy in the near future.

\section*{Acknowledgments}
I sincerely acknowledge Tapobrata Sarkar for helping me write this review article. I would also like to thank Pritam Banerjee for useful discussions. \tred{I sincerely thank the anonymous referees for their constructive suggestions.}


\end{document}